\documentclass{article}
\usepackage{amsmath,amssymb}
\usepackage{natbib}
\usepackage{graphicx,caption}

\newcommand{\dd}{\,\mathrm d}
\newcommand{\gm}{\gamma}
\newcommand{\kp}{\kappa}
\newcommand{\R}{\mathbb{R}}

\title{Globally intensity-reweighted estimators for $K$- and pair correlation functions}
\author{Thomas Shaw\thanks{Applied Physics Program, University of
Michigan, 1425 Randall Laboratory, 450 Church St, Ann Arbor, MI,
48109, USA},
Jesper M{\o}ller\thanks{Department of Mathematical Sciences, Aalborg
University, Skjernvej 4A, 9220 Aalborg {\O}, Denmark, \hspace*{1ex} Email:
\texttt{jm@math.aau.dk}},
and Rasmus Plenge Waagepetersen\thanks{Department of Mathematical Sciences, Aalborg University, Skjernvej 4A, 9220 Aalborg {\O}, Denmark} }
\begin{document}
\maketitle

\begin{abstract}
We introduce new estimators of the inhomogeneous $K$-function and the pair
correlation function of a spatial point process as well as the cross
$K$-function and the cross pair correlation function of a bivariate
spatial point process under the assumption of second-order
intensity-reweighted stationarity. These estimators rely on a `global'
normalization factor which depends on an aggregation of the intensity
function, whilst the existing estimators depend `locally' on the
intensity function at the individual observed points. The
  advantages of our new global estimators over the existing local
  estimators are demonstrated by theoretical considerations and a simulation study.

\end{abstract}

\footnotetext[0]{{\sl Acknowledgment.} The first author was 
supported by grants from the U.S. National Science Foundation (MCB1552439) and National
Institutes of Health (R01GM129347) and he thanks his advisor Dr.\ Sarah Veatch for her support.
The last two authors were supported by the `Danish Council for Independent Research | Natural Sciences' grant DFF -- 7014-00074
`Statistics for point processes in space and beyond', and by the `Centre for Stochastic Geometry and
Advanced Bioimaging' funded by grant 8721 from the Villum Foundation.}

\section{Introduction}

Functional summary statistics like the nearest-neighbour-, the empty space-,
and Ripley's $K$-function have a long history in statistics for spatial point
processes
\citep{MoellerWaagepetersenl2004,IllianStatistical2008,chiu2013stochastic}. For
many years the theory of these functional summary statistics was confined to
the case of stationary point processes with consequently constant intensity
functions. The paper \cite{BaddeleyNon2000} was therefore a big step forward
since it relaxed substantially the assumption of stationarity in case of the
$K$-function and the closely related pair correlation function.

\cite{BaddeleyNon2000} introduced the notion of second-order
intensity-reweighted stationarity ({\emph{soirs}}) for a spatial
point process. When the pair correlation function $g$ exists for the
point process, soirs is equivalent to that $g$ is translation
invariant.  However, the intensity function does not need to be
constant which is a great improvement compared to assuming
stationarity, see e.g.\ \cite{MollerModern2007}. When the point
process is soirs, \cite{BaddeleyNon2000} introduced a generalization of Ripley's $K$-function, the so-called inhomogeneous $K$-function which is based on the idea of intensity-reweighting the points of the spatial point process, and they
discussed its estimation. The inhomogeneous  $K$-function has found applications in
a very large number of applied papers and has also been generalized
e.g.\ to the case of space-time point processes
\citep{GabrielSecondorder2009} and to point processes on spheres
\citep{lawrence:etal:16,moeller:rubak:16}. Moreover, van Lieshout (2011)
\nocite{lieshout:11} used the idea of intensity-reweighting to generalize the so-called $J$-function to the case of inhomogeneous point processes.

A generic problem in spatial statistics,  when just one realization of
a spatial process is available, is to separate variation due to random
interactions from variation due to a non-constant intensity or mean
function. In general, if an informed choice of a  parsimonious
intensity function model is available for a point process, the
intensity function can be estimated consistently. Consistent
estimation of the inhomogeneous $K$-function is then also possible when the
consistent intensity function estimate is  used to reweight the point
process, see e.g.\  \cite{waagepetersen:guan:09} in case
  of regression models for the intensity function. When a parsimonious
  model is not available, one may resort to non-parametric kernel estimation of the intensity function as considered initially in \cite{BaddeleyNon2000}. However, kernel estimators are not consistent for the intensity function and they are strongly upwards biased when evaluated at the observed points. This implies strong bias of the resulting inhomogeneous $K$-function estimators when the kernel estimators are plugged in for the true intensity.

In this paper, we introduce a new approach to non-parametric
estimation of the (inhomogeneous) $K$ and $g$-functions for a spatial point process,
or of the cross $K$-function and the cross pair correlation for a
bivariate spatial point process, assuming soirs in both
cases.
This formalizes an approach that was used by \cite{stone:veatch:2017} to
estimate space-time cross pair correlation functions in live-cell single molecule localization
microscopy experiments with spatially varying localization probabilities.
In the univariate case, our new as well as the existing
estimators are given by a sum over all distinct points $x$ and $y$  from
an observed point pattern. For the new estimators, each term in
  the sum depends on an aggregation of the intensity function through a `global' normalization factor $\gamma(y-x)$ instead
  of depending `locally' on the intensity function at $x$ and at $y$
  as for the existing estimators (a similar remark applies in the
  bivariate case). Intuitively one may expect this to mitigate the
problem of using biased kernel estimators of the intensity function in
connection to non-parametric estimation of the $K$-function or pair
correlation function. Moreover, to reduce bias when using a
non-parametric kernel estimator of $\gamma$, we propose a `leave-out' modification
of our $\gamma$ estimator.
 Our simulation study shows that our
  new globally intensity reweighted estimators are superior to the
  existing local estimators in terms of bias and estimation variance regardless
  of whether the intensity function is estimated parametrically or non-parametrically.

The remainder of the paper is organized as
follows. Some background on spatial
  point processes and notational details are provided in Section~\ref{s:pre}. Section~\ref{s:K} introduces our global estimator
for the $K$-function or the cross $K$-function, discusses
modifications to account for isotropy, and compares with the existing
local estimators. Section~\ref{s:g} is similar but for our new
global estimator of the $g$-function or cross pair correlation
function. Section~\ref{s:bias} describes
sources of bias in the local and global estimators
when kernel estimators are
used, and modifications to reduce bias. In Section~\ref{s:sim}, the
global and local estimators of $K$ and $g$ are compared in a
simulation study. Possible extensions are discussed in
  Section~\ref{s:conrem}. Finally, Section~\ref{s:conclusion} contains some
  concluding remarks.

\section{Preliminaries}\label{s:pre}

We consider the usual setting for a spatial point process $X$ defined on
the $d$-dimensional Euclidean space $\mathbb{R}^d$, that is,
 $X$ is a random locally finite subset
of $\mathbb R^d$. This means that the number of points from $X$ falling in $A$,
denoted $N(A)$, is almost surely finite for any bounded subset $A$ of
$\mathbb{R}^d$.  For further details we refer to
\cite{MoellerWaagepetersenl2004}. In our examples, $d=2$.

For any integer $n\ge 1$, we say that $X$ has \emph{$n$-th order
  intensity function} $\rho^{(n)}:(\mathbb R^d)^n\mapsto[0,\infty)$ if
for any disjoint bounded Borel sets $A_1,\ldots,A_n\subset\mathbb R^d$,
\[\mathrm E\{N(A_1)\cdots N(A_n)\}=\int_{A_1}\cdots\int_{A_n}\rho^{(n)}(x_1,\ldots,x_n)\,\mathrm
  dx_1\,\cdots\mathrm dx_n<\infty.\]
By the so-called standard proof we obtain the $n$-th order Campbell's formula \citep[see e.g.][]{MoellerWaagepetersenl2004}: for any Borel function $k:(\mathbb R^d)^n\mapsto[0,\infty)$,
\[\mathrm E\sum^{\not=}_{x_1,\ldots,x_n\in X}k(x_1,\ldots,x_n)=\int\cdots\int k(x_1,\ldots,x_n)\rho^{(n)}(x_1,\ldots,x_n)\,\mathrm
  dx_1\,\cdots\mathrm dx_n,\] 
  which is finite if the left or right hand side is so. Here, $\not=$ over the summation sign means that $x_1,\ldots,x_n$ are pairwise distinct. 

Throughout this paper, we assume that $X$ has an \emph{intensity function}
$\rho$ and a translation invariant \emph{pair correlation function} $g$. This means that for all $x,y \in \mathbb{R}^d$, $\rho^{(1)}(x)=\rho(x)$ and $\rho^{(2)}(x,y)=\rho(x)\rho(y)g(x,y)$, where
$g(x,y)=g_0(x-y)$ with $g_0:\mathbb R^d\mapsto[0,\infty)$ a symmetric
Borel function.
If $\rho$ is constant we say that $X$ is
  (first-order) \emph{homogeneous}. In particular, if $X$ is stationary, that is, the distribution of $X$ is invariant under translations in $\mathbb R^d$, then $\rho$ is constant and $g$ is translation invariant.
  
Following  \cite{BaddeleyNon2000}, the translation invariance of $g$ implies that 
$X$ is second-order intensity reweighted stationary (soirs) and the \emph{inhomogeneous $K$-function} (or just $K$-function) is then given by
\[K(t):=\int_{\|h\|\le t}g_0(h)\,\mathrm dh,\qquad t\ge0.\]
This is Ripley's $K$-function when $X$ is stationary.

Suppose $X_1$ and $X_2$ are locally finite
point processes on $\mathbb R^d$ such that $X_i$ has intensity function $\rho_i$, $i=1,2$, and
$(X_1,X_2)$ has a translation invariant \emph{cross pair correlation function}
$g_{12}(x_1,x_2)=c(x_1-x_2)$ for all $x_1,x_2\in\mathbb R^d$.  That is, for bounded Borel sets
$A_1,A_2\subset\mathbb R^d$ and   $N_i(A_i)$ denoting the cardinality of
$X_i\cap A_i$, $i=1,2$, we have
\[\mathrm E\{N_1(A_1)N_2(A_2)\}=\int_{A_1}\int_{A_2} \rho_1(x_1)\rho_2(x_2)c(x_1-x_2)\,\mathrm dx_1\,\mathrm dx_2.\] 
Then the \emph{cross $K$-function} is defined by
\[K_{12}(t):=\int_{\|h\|\le t}c(h)\dd h,\qquad t\ge0.\]

In practice $X,X_1,X_2$ are observed within a
bounded window $W \subset \mathbb{R}^d$, and we use the following notation. The
translate of $W$ by $x\in\mathbb R^d$ is denoted $W_x:=\{w+x\,|\,w\in W\}$. For a Borel set
$A\subseteq\mathbb R^d$, $1[x\in A]$ denotes the indicator
function which is 1 if $x\in A$ and 0 otherwise. The Lebesgue measure
of $A$ (or area of $A$ when $d=2$) is denoted $|A|$, and $\|x\|$ is the
usual Euclidean length of $x\in\mathbb R^d$.

\section{Global and local intensity-reweighted estimators for $K$-functions}\label{s:K}

\subsection{The case of one spatial point process}

Considering the setting in Section~\ref{s:pre} for the spatial point process $X$, we
define
\begin{equation}\label{eq:denominator} \gamma(h):=\int_{W \cap W_{-h}} \rho(u) \rho(u + h)\,\mathrm d u,\qquad h\in\mathbb R^d.
\end{equation}
Clearly, $\gamma$ is symmetric, that is, $\gamma(h)=\gamma(-h)$ for all $h\in\mathbb R^d$. 
We assume that with probability 1, 
$\gamma(y-x)>0$ for all distinct $x,y\in X\cap W$.
Then, for $t \ge 0$, we can define
\begin{equation}\label{e:Khat-global} \hat K_{\text{global}}(t) := \sum^{\not=}_{x,y\in X\cap W} \frac{1[\|y-x \| \le t]}{\gamma(y-x)}. 
\end{equation}
If $\gamma(h)>0$ whenever $\|h\|\le t$, then $\hat K_{\text{global}}(t)$  
is an unbiased estimator of $K(t)$. 
This follows from the second-order Campbell's formula:
\begin{align*}  \mathrm E \hat K_{\text{global}}(t)& = \int \int \frac{1[x \in W,y \in W, \|y-x\| \le t]}{\gamma(y-x)}\rho(x)\rho(y) g_0(y-x) \dd x \dd y\\
&= \int \int \frac{1[x \in W \cap W_{-h}, \|h\| \le t]}{\gamma(h)}\rho(x)\rho(x+h) g_0(h) \dd x \dd h \\
&= \int_{\|h\| \le t}  \frac{\gamma(h)}{\gamma(h)} 
g_0(h)\dd h = K(t) .
\end{align*}

We call $\hat K_{\text{global}}$ the \emph{global estimator} since it contrasts with one of the estimators suggested in \cite{BaddeleyNon2000}: assuming that almost surely $|W\cap W_{y-x}|>0$ for distinct $x,y\in X\cap W$,   
\begin{equation}\label{e:Khat-local} 
\hat K_{\text{local}}(t) := \sum^{\not=}_{x,y\in X\cap W} \frac{1[\|y-x \| \le t]}{\rho(x)\rho(y)|W\cap W_{y-x}|}, 
\end{equation}
which we
refer to as the \emph{local estimator}. Note that $\hat K_{\text{local}}(t)$ 
is also an unbiased estimator of $K(t)$ provided $|W\cap W_h|>0$ for $\|h\|\le t$. In the homogeneous case,
\[ \gamma(h)= \rho^2 |W \cap W_{-h}|, \]
whereby $\hat K_{\text{global}}=\hat K_{\text{local}}$, and in the stationary case, these estimators coincide with the \cite{OhserStoyan1981} translation estimator.

In practice $\rho$ and hence $\gamma$ must be replaced by
estimates. Estimators of $\rho$ and $\gamma$ and the bias of these
estimators are discussed in Section~\ref{s:bias}. 

\subsubsection{Modifications to account for isotropy} \label{s:isoK}

In addition to soirs, it is
frequently assumed that the pair correlation function is isotropic 
meaning that $g_0(h)=g_1(\|h\|)$ for some
Borel function $g_1:[0,\infty)\mapsto[0,\infty)$.
We benefit from this
  by integrating over the sphere: for $r > 0$, define
\begin{equation}
\gamma^\mathrm{iso}(r) := \int_{\mathbb{S}^{d-1}} \gamma(rs) \dd \nu_{d-1}(s)\big/\varsigma_d, \label{e:gammaiso}
\end{equation}
where $\mathbb S^{d-1}=\{s\in \mathbb R^d\,|\,\|s\|=1\}$ denotes the $(d-1)$-dimensional
unit-sphere, $\nu_{d-1}$ is the $(d-1)$-dimensional surface measure on $\mathbb{S}^d$, and 
$\varsigma_d = 2\pi^{d/2}/\Gamma(d/2)$ is
the surface area of the unit sphere $\mathbb{S}^{d-1}$. Thus $\gamma^\mathrm{iso}(r)$ is the mean value of $\gamma(H)$ when $H$ is a uniformly distributed point on the $(d-1)$-dimensional sphere of radius $r$ and center at the origin.

Assuming that almost surely $\gamma^\mathrm{iso}(\lVert y - x \rVert)>0$ for
distinct $x,y\in X\cap W$, this naturally leads to another
global estimator for $K$ when
the pair correlation function is isotropic, namely
\begin{equation}
\hat K_\mathrm{global}^\mathrm{iso}(t) :=
\sum_{x,y\in X \cap W}^{\neq}
\frac{1\left[\lVert y-x \rVert \le t\right]}{\gamma^\mathrm{iso}(\lVert y - x \rVert)}.
\label{e:isoKglobal}
\end{equation}
That $\hat K_\mathrm{global}^\mathrm{iso}$ is unbiased follows from a similar
derivation as for $\hat K_\mathrm{global}$: for any $t\ge0$ such that $\gamma^\mathrm{iso}(r)>0$ whenever $r\le t$,
\begin{align}
\mathrm{E} \hat K_\mathrm{global}^\mathrm{iso}(t)
&= \int_{\lVert h \rVert \le t}
\frac{\gamma(h)}{\gamma^\mathrm{iso}(\lVert h \rVert)}g_0(h) \dd h \nonumber \\
    &= \int_0^t g_1(r) r^{d-1}\int_{\mathbb S^{d-1}} \frac{\gamma(rs)}{\gamma^\mathrm{iso}(r)} \dd \nu_{d-1}(s) \dd r \label{e:topolarK}\\
&= \int_0^t g_1(r) {\varsigma_d}{r^{d-1}} \dd r \nonumber \\
&= \int_{\lVert h \rVert \le t} g_1(\lVert h \rVert ) \dd h = K(t), \label{e:frompolarK}
\end{align} 
where \eqref{e:topolarK} and \eqref{e:frompolarK} employ changes of variables
to and from polar coordinates, respectively.

When $X$ is homogeneous, \eqref{e:isoKglobal} coincides with the
\cite{OhserStoyan1981} isotropic estimator. A local estimator of this
form can also be defined:
\begin{equation}
\hat K_\mathrm{local}^\mathrm{iso}(t)
:= \sum_{x,y\in X \cap W} \frac{1[\lVert y-x \rVert \le t]}{\rho(x)\rho(y) a_W(\lVert y-x\rVert)},
\end{equation}
where
\begin{equation}
a_W(r) = \int_{\mathbb S^{d-1}} | W \cap W_{-rs} | \dd \nu_{d-1}(s) \big/\varsigma_d
\end{equation}
is an isotropized edge correction factor, and where it is assumed that almost surely $a_W(\lVert y-x\rVert)>0$ for distinct $x,y\in X\cap W$. The local estimator is unbiased when $a_W(r)>0$ for $r\le t$.

\subsubsection{Comparison of local and global estimators}\label{s:comparevar}

The global and local estimators \eqref{e:Khat-global} and
\eqref{e:Khat-local} differ in the relative weighting of distinct points
$x,y\in X\cap W$. Namely, $\hat K_{\mathrm{local}}$ weights pairs
$x,y$ from low-density areas more strongly than those from
high-density areas, whilst for $\hat K_{\mathrm{global}}$, the
  weight only depends on the difference $y-x$.
Theoretical expressions for the variances of the global and local
$K$-function estimators are very complicated, not least when the
intensity function is replaced by an estimate. This makes it difficult
to make a general theoretical comparison of the estimators in terms of
their variances. However, under some simplifying assumptions insight can be gained as explained in the following.

  Consider a quadratic observation window $W$ of sidelength $nm$. Then $W$ is a disjoint union of $n^2$ quadrats $W_1,\ldots,W_{n^2}$ each of sidelength $m$. Assume that the intensity function is constant and equal to $\rho_i$ within each $W_i$, with $\rho$ naturally estimated by $\hat\rho(u)=\hat \rho_i=N(W_i)/m^2$ for $u \in W_i$. For fixed $t$ and large $m$, when $\rho$ is replaced by its estimator $\hat\rho$,  we can now approximate the local estimator: 
  \begin{align*} \hat K_{\text{local}}(t) = & \sum_{u,v \in X\cap W}^{\neq} \frac{1[\|u-v\| \le t]}{\hat \rho(u) \hat \rho(v) |W \cap W_{u-v}|}  \simeq \sum_{i=1}^{n^2} \sum_{u,v \in X\cap W_i}^{\neq} \frac{1[\|u-v\| \le t]}{ \hat \rho_i^2 |W \cap W_{u-v}|}\\
    \simeq & \sum_{i=1}^{n^2} \sum_{u,v \in X\cap W_i}^{\neq} \frac{1[\|u-v\| \le t]}{ \hat \rho_i^2 |W_i \cap (W_i)_{u-v}|n^2}
    = \frac{1}{n^2} \sum_{i=1}^{n^2} \hat K_{i,\text{local}}(t).
  \end{align*}
where $\hat K_{i,\text{local}}$ is the local estimator based on $X \cap
W_i$. {We use here $\simeq$ in a rather loose sense, meaning
  that asymptotically, as $m$ tends to infinity, the difference
  between the two quantities on each side of $\simeq$ tends to zero
  in a suitable sense (e.g.\ in mean square) under appropriate
  regularity conditions.} The first approximation above follows because contributions
from $u \in X_i$ and $v \in X_j$, $i \neq j$, are negligible
for fixed $t$ and $m$ large, and  the second approximation
  is justified  since for $\|h\| \le t$, $|W|/|W \cap W_{h}|$ and
  $|W_i|/|W_i \cap (W_i)_{h}|$ will tend to 1 as $m$ increases.  
Following similar steps, we obtain for the global estimator,
  \[ \hat K_{\text{global}}(t) \simeq \sum_{i=1}^{n^2} \hat
    K_{i,\text{local}}(t) \frac{\hat \rho_i^2}{\sum_{l=1}^{n^2} \hat
      \rho_l^2}. \]

Suppose $X$ is a Poisson process. Note that $\hat K_{\text{local}}(t)$ is an equally weighted average of the
$\hat K_{i,\text{local}}(t)$, but since the $\hat K_{i,\text{local}}(t)$ are independent, the optimal weighted average is obtained
with weights inversely proportional to the variances of the
$\hat K_{i,\text{local}}(t)$. For large $m$, the variance of
 $\hat K_{i,\text{local}}(t)$ is well approximated by $2\pi
 t^2/(\rho_i^2 m^2)$ \citep{ripley:88,lang:marcon:13} and the
 optimal weights $w_i$ are thus proportional to $\rho_i^2$. Our global estimator is obtained from the optimal weighted average by
replacing the optimal weights by natural consistent
estimates. Hence one may anticipate that the global estimator has
smaller variance than the local estimator. In a small-scale simulation
study this was indeed the case, and the global estimator with (random)
weights proportional to $\hat \rho_i^2$ even had slightly smaller
variance than when the optimal fixed weights $w_i \propto \rho_i^2$ were used.
  
\subsection{The case of two spatial point processes}

For two spatial point processes $X_1$ and $X_2$ {observed on the same observation window $W$ (cf.\ Section~\ref{s:pre})}, we
define the following
global estimator for the cross $K$-function: for
$t\ge0$,
\begin{equation} \label{e:K12hat-global}
 \hat K_{12,\text{global}}(t)
 := \sum_{x \in X_1 \cap W,y\in X_2 \cap W} \frac{1[\|y-x \| \le t]}{\gamma_{12}(y-x)} 
\end{equation} 
where
\[ \gamma_{12}(h):= \int_{W \cap W_{-h}} \rho_1(u) \rho_2(u + h)\,\mathrm d u\]
and it assumed that almost surely $\gamma_{12}(y-x)>0$ for $x \in X_1 \cap W$ and $y\in X_2 \cap W$.
It is straightforwardly verified that $\hat K_{12,\text{global}}(t)$ is unbiased for any $t\ge0$ such that $\gamma_{12}(h)>0$ whenever $\|h\|\le t$.

The corresponding local estimator is
\begin{equation}\label{e:K12hat-local}
\hat K_{12,\text{local}}(t) := \sum_{x \in X_1 \cap W,y\in X_2 \cap W} 
\frac{1[\|y-x \| \le t]}{\rho_1(x)\rho_2(y)|W\cap W_{y-x}|},
\end{equation}
assuming that almost surely $|W\cap W_{y-x}|>0$ for $x \in X_1 \cap W$ and $y\in X_2 \cap W$. The local estimator is unbiased when 
$|W\cap W_{h}|>0$ for $\|h\|\le t$.

Interchanging $X_1$ and $X_2$ does not affect
  \eqref{e:K12hat-global}: $\hat K_{12,\text{global}}(t)=\hat
  K_{21,\text{global}}(t)$ when $\hat K_{21,\text{global}}(t)$ is defined as in
\eqref{e:K12hat-global}  with $\gamma_{12}$ replaced by 
\[ \gamma_{21}(h):= \int_{W \cap W_{-h}} \rho_1(u+h) \rho_2(u) \dd u.\]
This follows since by a change of variable, $\gamma_{12}$ is
symmetric, $\gamma_{21}(h) = \gamma_{12}(-h)=\gamma_{12}(h)$.

When the cross pair correlation function $c(h)$ is also isotropic,
additional unbiased estimators of $K_{12}$ are readily obtained in the same way
as for the one point process case. Thus, defining
\begin{equation}\label{e:gamma12-iso}
\gamma_{12}^\text{iso}(r) := \int_{\mathbb{S}^{d-1}} \gamma_{12}(rs) \dd \nu_{d-1}(s) \big/\varsigma_d,\qquad r\ge0,
\end{equation}
and assuming that almost surely $\gamma_{12}^\text{iso}(\|y-x\|)>0$ for $x\in X_1 \cap W$ and $y\in X_2$,
we define an isotropic global estimator by
\begin{equation}\label{e:K12hat-global-iso}
\hat K_{12,\text{global}}^\text{iso}(t) := \sum_{x\in X_1 \cap W, y\in X_2 \cap W}^{\not=}
\frac{1[\|y-x\| \le t]}{\gamma_{12}^\text{iso}(\|y-x\|)}.
\end{equation}
This is easily seen to be unbiased when $\gamma_{12}^\text{iso}(r)>0$ for $r\le t$. 
Finally, the isotropic local estimator is
\begin{equation}\label{e:K12hat-local-iso}
\hat K_{12,\text{local}}^\text{iso}(t) := \sum_{x\in X_1 \cap W, y\in X_2 \cap W}^{\not=}
\frac{1[\|y-x\| \le t]}{\rho_1(x)\rho_2(y)a_W(\|y-x\|)},
\end{equation}
with $a_W(r)$ as defined in Section~\ref{s:isoK}, and it becomes unbiased if $a_W(r)>0$ for $r\le t$.

\section{Global and local intensity-reweighted estimators for pair correlation functions}\label{s:g}

\subsection{The case of one spatial point process}

Considering again the setting in Section~\ref{s:pre} for the spatial point
process $X$, this section introduces global and local estimators for the translation invariant pair correlation function given by $g_0$. Note that it may be easier to interpret $g_0$ than $K$, but non-parametric
kernel estimation of $g_0$ involves the choice of a bandwidth.

  Let $\kappa_b:\mathbb R^d\mapsto[0,\infty)$ be
a (normalized) kernel with bandwidth
$b>0$, that is, $\kappa_b(h)=\kappa_1(h/b)/b^d$ for $h\in\mathbb R^d$,
where $\kappa_1$ is a probability density function. We assume that $\kappa_1$
has  support centered in the origin and contained in $[-k,k]^d$ for some $k>0$;
 e.g.\ $\kappa_1$ could be a standard $d$-dimensional normal density
 truncated to $[-k,k]^d$ (this choice is convenient when $W$ is
 rectangular with sides parallel to the usual axes in $\mathbb
 R^d$). {Note that the bounded support of $\kappa_b$
   shrinks to $\{0\}$ when $b$ tends to zero}. Then,
for $h\in\mathbb R^d$,
\begin{align}
\mathrm{E} \sum^{\not=}_{x,y\in X\cap W} &\kappa_b(h -
                                           (y-x))
                                           \nonumber \\ 
&= \int_W \int_W \kappa_b(h - (y-x)) \rho(x)\rho(y)g_0(y-x)\,\mathrm dx\,\mathrm dy \label{e:Cam}\\
&=  \int_W \left\{\int_{W_{-h-x}}\kappa_b(-z) \rho(x)\rho(x + h + z)g_0(h+z)\,\mathrm dz\right\}\,\mathrm dx \nonumber \\
&\simeq g_0(h) \int_W \rho(x)\left\{\int_{W_{-h-x}}  \kappa_b(-z)  \rho(x + h + z)\,\mathrm dz\right\}\,\mathrm dx \label{approx1}\\
&\simeq g_0(h) \gamma(h) \label{approx2}
\end{align}
where $\gamma(h)$ is defined in \eqref{eq:denominator}.
Here, \eqref{e:Cam} follows from the second-order Campbell's formula and {$\simeq$ in  \eqref{approx1} and
\eqref{approx2} means that the difference between the quantities on
each side of $\simeq$ converges to zero as the bandwidth $b$ tends to
zero, under appropriate {continuity} conditions on $\rho(\cdot)$ and $g_0(\cdot)$. 
The expression \eqref{approx1} is expected to be more accurate but
\eqref{approx2} is simpler to compute}.

From \eqref{approx2}
we conclude that $g_0(h)$ can be estimated by the following \emph{global estimator},
\begin{equation*}
\hat g_{\mathrm{global}}(h) := { \sum^{\neq}_{x,y\in X\cap W} \kappa_b(h - (y-x))}\big/
 \gamma(h),
\end{equation*}
provided $\gamma(h)>0$.
This contrasts with the \emph{local estimator}
\begin{equation*}
\hat g_{\mathrm{local}}(h) := \sum_{x,y \in X\cap W}^{\neq} {\kappa_b(h - (y-x))}\big/
\left\{{\rho(x) \rho(y) ~ | W \cap W_{x-y}|}\right\},
\end{equation*}
which is analogous to the estimator suggested in
\cite{BaddeleyNon2000} for an isotropic pair correlation function, see also
Section~\ref{s:three}.

\subsubsection{Modifications to account for isotropy}\label{s:three}

For isotropic point processes as defined in Section \ref{s:isoK}, the global
pair correlation function estimator may be modified to estimate the isotropic
pair correlation function given by $g_1$: for $r > 0$ such that $\gamma^\mathrm{iso}(r)>0$, define
\begin{equation}
\hat g_{\mathrm{global}}^\mathrm{iso}(r):= \frac{1}{\varsigma_d r^{d-1}}
   \sum_{x, y \in X\cap W}^{\neq} \tilde \kappa_b(r - \lVert x - y \rVert)  
 \big/ \gamma^\mathrm{iso}(r) ,\label{e:iso}
\end{equation}
where for $b>0$, $\tilde \kappa_b(t)=\tilde \kappa_1(t/b)/b$, $t \in
\R$, for a probability density $\tilde \kp_1 :\mathbb{R} \mapsto
[0,\infty)$ with support centered at 0
and contained in the interval $[-k,k]$ for some constant $k >0$, and where $\gamma^\mathrm{iso}(r)$ is defined in \eqref{e:gammaiso}.
This definition is motivated by the following derivation:
\begin{align}
&\mathrm{E} \sum_{x,y \in X\cap W}^{\neq} \tilde \kappa_b(r - \lVert y - x \rVert) \nonumber\\
&= \int_W \int_W \tilde \kappa_b (r - \lVert y - x \rVert)
\rho(x) \rho(y) g_1(\lVert y - x \rVert)\,\mathrm dy\, \mathrm dx \label{e:a}\\
&= \int_W \left\{\int_0^\infty \tilde \kappa_b (r - \xi) g_1(\xi)\xi^{d-1}
\int_{\mathbb S^{d-1}} \rho(x) \rho(x + \xi s) 1[x + \xi s \in W]
                                                                                  \,\mathrm d\nu_{d-1}(s)\,\mathrm d\xi\right\}\, \mathrm dx \label{e:b} \\
&\simeq 
g_1(r) \varsigma_d \gamma^\mathrm{iso}(r) r^{d-1} \int_0^\infty \tilde\kappa_b (r - \xi)\,\mathrm d\xi \label{e:c}\\
& \simeq
g_1(r) \varsigma_d \gamma^\mathrm{iso}(r) r^{d-1},\label{e:d}
\end{align}
using the second-order Cambell formula in \eqref{e:a}, a `shift to polar
coordinates' in \eqref{e:b}, the assumption that $b$ is small in \eqref{e:c},
and that the kernel is a probability density function in \eqref{e:d}. Note
regarding \eqref{e:d} that 
\[ \int_0^\infty \tilde\kappa_b (r - \xi) \dd \xi = \int_{-\infty}^r \tilde\kappa_b (\xi) \dd \xi \]
which is not $1$ in general. Since $\tilde\kappa_b (\xi)=0$ for $\xi\not\in[-bk,bk]$, the integral is 1 if $bk<r$. From \eqref{e:d} we obtain \eqref{e:iso}.

In the isotropic case the most commonly used local estimators \citep{BaddeleyNon2000} are
\[  \hat g_{\mathrm{local}}^{\mathrm{iso}}(r) = \frac{1}{\varsigma_d r^{d-1}}
  \sum_{x,y \in X \cap W}^{\neq} 
  \frac{ \tilde \kappa_{b}(r - \|y-x\|)}{ \rho(x) \rho(y)|W \cap W_{x-y}|} \]
and
\[ \tilde g_{\mathrm{local}}^{\mathrm{iso}}(r) = \frac{1}{\varsigma_d }
  \sum_{x,y \in X \cap W}^{\neq} 
  \frac{ \tilde \kappa_{b}(r - \|y-x\|)}{ \rho(x) \rho(y)|W \cap W_{x-y}|\|y-x\|^{d-1}},
\]
assuming that almost surely $|W \cap W_{x-y}|>0$ for distinct $x,y \in X \cap W$.
These estimators suffer from strong positive respectively negative
bias for values of $r$ close to 0.

\subsection{Two point processes}

A similar derivation is possible for the cross pair correlation function of a
bivariate point process  $(X_1,X_2)$, yielding similar global and local estimators of $c(h)$: for $\gamma_{12}(h)>0$,
\begin{equation*}
\hat c_{\mathrm{global}}(h) := {\sum_{x\in X_1\cap W, y \in X_2\cap W} \kappa_b (h - (y - x))}\big/
\gamma_{12}(h);
\end{equation*}
for $\gamma_{12}^\mathrm{iso}(r)>0$,
\begin{equation*}
\hat c_{\mathrm{global}}^\mathrm{iso}(r) := \frac{1}{\varsigma_d r^{d-1}}{\sum_{x\in X_1\cap W, y \in X_2\cap W} \tilde \kappa_b (r - \|y - x\|)}\big/
\gamma_{12}^\mathrm{iso}(r);
\end{equation*}
and for $| W \cap W_{x-y} |>0$ almost surely when $x\in X_1\cap W$ and $y\in X_2\cap W$,
\begin{equation*}
\hat c_{\mathrm{local}}(h) = \sum_{x\in X_1\cap W, y \in X_2\cap W} {\kappa_b(h - (y - x))}/
\left\{{\rho_1(x) \rho_2(y)~ | W \cap W_{x-y} | }\right\}
\end{equation*}
and
\begin{equation*}
\hat c_{\mathrm{local}}^\mathrm{iso}(r) = \frac{1}{\varsigma_d r^{d-1}}{\sum_{x\in X_1\cap W, y \in X_2\cap W}
\frac{\tilde \kappa_b (r - \|y - x\|)}{\rho_1(x)\rho_2(y)|W\cap W_{x-y}|}}.
\end{equation*}

Also
an intermediate estimator is possible,
with the intensity weighting for one of the processes applied locally, and the
other applied globally: with $X_1$, $X_2$, and $\kappa_b$ as above, we have
\begin{align*}
\mathrm{E} \sum_{x \in X_1\cap W, y \in X_2\cap W}& \frac{\kappa_b(h-(y-x))}{\rho_2(y)}
\\
&= \int_W \int_W \kappa_b(h - (y-x)) c(y-x) \rho_1(x)\,\mathrm dx\,\mathrm dy \\
&= \int_W\int_{W_{-x-h}} \kappa_b(-z)c(h+z) \rho_1(x) \dd z \dd x \\
&\simeq c(h) \int_{W \cap  W_{-h}} \rho_1(x) \dd x 
\end{align*}
for a small bandwidth $b>0$, which suggests the partially-reweighted estimator
\begin{equation*}
\hat c_{\mathrm{partial}}(h) :=  \sum_{x \in X_1\cap W, y \in X_2\cap W} \frac{
\kappa_b(h - (y-x))}{\rho_2(y)\int_{W \cap  W_{-h}} \rho_1(x) \dd x}, \label{e:partial}
\end{equation*}
provided $\int_{W \cap  W_{-h}} \rho_1(x) \dd x>0$.
This estimator may be useful when $\rho_2$ is much easier to estimate than
$\rho_1$, e.g.\ when $X_2$ is homogeneous.

\section{Sources of bias when $\rho$ is estimated} \label{s:bias}

All of the estimators of $K(t)$, $K_{12}(t)$, $g_0(h)$, and $g_1(r)$ discussed above are unbiased (at least when $t,\|h\|,r$ are  sufficiently small) when the
true intensity function $\rho$ is used to compute the weight
functions $\rho(x)\rho(y)$ in the local estimators or $\gamma$, $\gamma^\mathrm{iso}$, $\gamma_{12}$, or $\gamma_{12}^\mathrm{iso}$
in the global estimators. However, in most applications $\rho$ is not
known, and must be replaced by an estimate.
 When the source of inhomogeneity is
well understood, it is recommended to fit a model with an appropriate parametric intensity function and use it as the estimate, cf.\ \cite{BaddeleyNon2000} and \cite{waagepetersen:guan:09}. 

In the absence of such a
model, the most common alternative is a kernel estimator
\begin{equation}
\hat \rho(x) := \sum_{y \in X\cap W} \frac{\kappa_\sigma(y - x) }{ w_W(x;y)}
\label{e:kernelrho}
\end{equation}
where $\kappa_\sigma$ is a symmetric kernel on $\mathbb{R}^d$ with
bandwidth  $\sigma>0$, and where  $w_W(x;y)$ is an appropriate edge correction weight. We
take the  standard choice from \cite{DiggleKernel1985},
\begin{equation*}
w_W(x;y) = \int_W \kappa_\sigma(u - x) \dd u,
\end{equation*}
 see also
\cite{vanLieshoutEstimation2012} (other types of edge corrections
may depend on both $x$ and $y$ which is why we write $w_W(x;y)$
although the weight here only depends on $x$.)

In the following we discuss estimators for  $\rho$ and $\gamma$ with
particular focus on the implications of estimation bias when kernel estimators are used to replace the true $\gamma$ or $\rho$ in the global
and local estimators.

\subsection{Bias of local estimators with estimated
  $\rho$} \label{s:local-bias}

We start by considering a single spatial point process $X$.
For each point pair $x,y\in X$ ($x\not=y$), the corresponding term in the local $K$- and pair correlation function estimators is normalized
by the product $\rho(x)\rho(y)$. While an exact expression
for the bias of the estimators with estimated $\rho$ is not analytically
tractable, we can understand major sources of bias by considering the
expression $1/(\hat \rho(x) \hat\rho(y))$, which appears in each of the local
estimators.

First, following \cite{BaddeleyNon2000}, we note that $\hat\rho$ as defined in
\eqref{e:kernelrho} is subject to bias when evaluated at the points of $X$, and
that a `leave-one-out' kernel estimator
given by
\begin{equation}
\bar \rho(x) := \sum_{y \in (X\cap W)\setminus \{x\}} \frac{\kappa_\sigma(y - x) }{ w_{W}(x;y)},\qquad x\in W,
\label{e:leaveoutkernel}
\end{equation}
is a better choice, with reduced bias in most cases.

Second, we note that \[\text{E}(1/\bar\rho(x)) > 1/\text{E}(\bar \rho(x))\] (if
$\text{E}(1/\bar\rho(x))$ exists; in some cases it may be infinite). This follows from
Jensen's inequality, since $x \mapsto 1/x$ is strictly convex for $x>0$. In addition, note
that the leading contribution to $\text{E}(1/\bar\rho(x)) - 1/\text{E}(\bar\rho(x))$ is
proportional to $\text{Var}\bar\rho(x)$ \citep{liao:berg:19}. This discrepancy leads to
a strong positive bias of the local $K$- and pair correlation function
estimators, especially at large $\|y -x \|$, where $1/\bar \rho(x)$ and
$1/\bar \rho(y)$ are almost independent. This effect becomes more pronounced
for smaller $\sigma$, since $\text{Var}\bar\rho(x)$ typically increases as $\sigma$ decreases.

Third, we note that
for distinct points $x,y\in W$ that
are close compared to the bandwidth $\sigma$, the covariance of $\bar\rho(x)$
and $\bar\rho(y)$ leads to bias. For the local (and global) estimators, we
  consider sums over distinct $x,y\in X \cap W$, which leads us to condition on $x,y\in X$ as follows \citep[for details, see][]{CoeurjollyPalm2017}. By $X$ conditioned on distinct points $x,y\in X$ with $\rho^{(2)}(x,y)>0$,
we mean that $X$ is equal to $X_{xy}\cup\{x,y\}$ in distribution, where $X_{xy}$ follows the second-order reduced Palm distribution of $X$ at $x,y$:
\[\mathrm P(X\in F\mid x,y\in X)=\mathrm P(X_{xy}\cup\{x,y\}\in F).\]
Assuming $X$ has $n$-th order joint intensity functions $\rho^{(n)}$
for $n\le4$,
 $X_{xy}$ has intensity function $\rho_{xy}(u)=\rho^{(3)}(x,y,u)/\rho^{(2)}(x,y)$ and second order joint intensity function $\rho^{(2)}_{xy}(u,v)=\rho^{(4)}(x,y,u,v)/\rho^{(2)}(x,y)$.
Now, for distinct $x,y\in W$ with $\rho^{(2)}(x,y)>0$, neglecting the edge
correction in \eqref{e:leaveoutkernel} for simplicity, we obtain the following by the first and second-order Campbell's formulas for $X_{xy}$ and using that $\kappa_\sigma$ is symmetric:
\begin{align}
    \text{E}& \left[ \bar \rho(x) \bar \rho(y) \,\big|\, x,y\in X\cap W \right]
= \text{E} \left\{\sum_{u \in (X_{xy}\cap W)\cup \{y\}} \kappa_\sigma(x - u)
\sum_{v \in (X_{xy}\cap W)\cup\{x\} } \kappa_\sigma(y - v) \right\} \nonumber\\
&=\text{E}\sum^{\not=}_{u,v\in X_{xy}\cap W}\kappa_\sigma(x-u)\kappa_\sigma(y-u)+\text{E}\sum_{u\in X_{xy}\cap W}\kappa_\sigma(x-u)\kappa_\sigma(y-u)\nonumber\\
&+\kappa_\sigma(x-y)\kappa_\sigma(y-x)\\
&+\kappa_\sigma(x-y)\text{E}\sum_{v\in X_{xy}\cap W}\kappa_\sigma(y-v)+\kappa_\sigma(y-x)\text{E}\sum_{u\in X_{xy}\cap W}\kappa_\sigma(x-u)\nonumber\\
&=
\int_W\int_W \kappa_\sigma(x-u)\kappa_\sigma(y-v) \frac{\rho^{(4)}(x,y,u,v)}{\rho^{(2)}(x,y)} \dd u \dd v \label{e:bbbbb} \\
&+ \int_W \kappa_\sigma(x-u)\kappa_\sigma(y-u)\frac{\rho^{(3)}(x,y,u)}{\rho^{(2)}(x,y)} \dd u
\label{e:barrhobarrho}\\
&+ \kappa_\sigma(x-y)^2 + \kappa_\sigma(x-y)\int_W \left\{\kappa_\sigma(x-u)
+ \kappa_\sigma(y-u)\right\}\frac{\rho^{(3)}(x,y,u)}{\rho^{(2)}(x,y)} \dd u\label{e:ccccc}.
\end{align}

If $X$ is a Poisson process, then $X$ and $X_{xy}$ are identically distributed, and so
the term in \eqref{e:bbbbb}  simplifies to $\text{E}\bar
\rho(x)\text{E}\bar\rho(y)$, which differs from $\rho(x)\rho(y)$ only by
the inherent bias of the kernel estimators.
In general, the joint intensity
$\rho^{(4)}(x,y,u,v)$ in the integrand of that term represents the additional
covariance of $\bar\rho(x)$ and $\bar\rho(y)$ due to interactions between the
points of the process, and induces further bias. For example, this bias will
tend to overestimate $\rho(x)\rho(y)$ for clustered processes, and lead to
an underestimate of $K$, $g_0$, and $g_1$.
The terms in \eqref{e:barrhobarrho} and \eqref{e:ccccc} are
non-negative, and in particular the
term in \eqref{e:barrhobarrho}  
can be large when $x$
and $y$ are close together compared to $\sigma$. This positive bias leads to substantial negative bias at short
distances of the local estimators of $K$, $g_0$, and $g_1$.

In comparison, the conditional expectation
$\text{E}\{\hat \rho(x) \hat \rho(y) \mid x,y\in X\}$ would have additional positive terms
depending on $\kappa(0)$.  
In the two point process case, the  relevant conditional expectation $\mathrm{E}\{ \bar \rho_1(x) \bar \rho_2(y) \mid x \in X_1, y \in X_2\}$ has an expression (of which we omit the details) analogous to \eqref{e:barrhobarrho}. However, since $X_1$ and $X_2$ are assumed to have a cross pair correlation function, almost surely
$u=v$ does not occur for $u \in X_1$ and $v \in X_2$, so no term analogous to the second term in \eqref{e:barrhobarrho} occurs in $\mathrm{E}\{  \bar \rho_1(x) \bar \rho_2(y) \mid x \in X_1, y \in X_2\}$. This reduces the bias problem in the two point process case compared to the single point process case.

For distinct $x,y\in W$ with $\rho^{(2)}(x,y)>0$, a superior estimator for $\rho(x)\rho(y)$ might be given by
\begin{equation}
\overline{\rho(x)\rho(y)} := \sum_{u,v \in X\cap W\setminus\{x,y\}}^{\neq}
\frac{\kappa(x-u)\kappa(y-v)}{w_W(x;u)w_W(y;v)}.
\label{e:barrhorho-leaveout}
\end{equation}
Then the terms in \eqref{e:barrhobarrho} and \eqref{e:ccccc} are
  avoided, since
\[
\text{E} \{ \overline{\rho(x)\rho(y)} \mid x,y \in X\cap W \} 
= \int_W\int_W \frac{\kappa(x-u)\kappa(y-v)}{w_W(x;u)w_W(y;v)}
\frac{\rho^{(4)}(x,y,u,v)}{\rho^{(2)}(x,y)} \dd u \dd v.
\]
We do not investigate this idea further in the current work.

\subsection{Bias of global estimators with estimated
  $\gamma$}\label{s:f-est}

Given the kernel estimate in \eqref{e:kernelrho} an immediate
  estimator of $\gm(h)$, $h \in \mathbb{R}^d$, is 
\begin{equation}
\hat \gamma(h) :=  \int_{W\cap W_{-h}}\hat \rho(z) \hat \rho(z + h) \dd z.
\label{e:kernelgamma}
\end{equation}
To understand properties of this estimator we evaluate its expected
value. We start with the simplest case
   where $h$ is a fixed vector in $\mathbb{R}^d$. This case is relevant for the global estimator of the pair correlation function. We
   return in the end of this section to the case where $h$ is an
   observed difference $h=y-x$ for distinct
   $x,y \in X$, which occurs for the global estimator of the $K$-function. 
   
   Neglecting edge corrections for simplicity, we get
\begin{align}
\mathrm E \hat \gamma(h)
&= \int_{W \cap W_{-h}} \int_W \kappa_\sigma(z-u) \rho(u) 
\int_W \kappa_\sigma(z+h - v) \rho(v)
g_0(u - v) \dd v \dd u \dd z\label{e:ddddd}\\
&+ \int_{W \cap W_{-h}} \int_W \kappa_\sigma(z - u) \kappa_\sigma(z+h - u) \rho(u)  \dd u \dd z.\label{e:eeeee}
\end{align}
The two resulting terms are analogous to the terms in \eqref{e:bbbbb} and \eqref{e:barrhobarrho}.

When $g_0 = 1$ as for a Poisson process, the term in the right hand side of \eqref{e:ddddd}
simplifies to
\[ \int_{W\cap W_{-h}} \mathrm{E}\hat \rho(x) \mathrm{E} \hat \rho(x + h) \dd x . \]
This differs from $\gamma(h)$  due to the inherent bias of
the kernel estimators which depends on the spatial structure of the intensity
function: $\mathrm{E}\hat\rho(x) - \rho(x)$ becomes large when $\sigma$ is large
compared to the length scale of spatial variation of $\rho(x)$.
On the other hand,
when $g_0\neq 1$, the term in the right hand side of \eqref{e:ddddd} 
includes an additional bias due to
the interaction between points. For example, this bias will tend to overestimate $\gamma$ for clustered
processes, and therefore lead to an underestimate of $K$ or the pair correlation function.
{This interaction bias is most pronounced when $\sigma$ is small. In particular, as
$\sigma \to 0$, this term approaches $g_0(y-x)\gamma(y-x)$, so that e.g.
$\mathrm{E}\hat g_\mathrm{global}(h)\to 1$ for all $h\in \mathbb{R}^d$. However, in the typical case where the strength of
pairwise interactions decreases with distance, increasing $\sigma$ reduces bias due to
interactions. Therefore, it is important to choose $\sigma$ to be larger than the length-scale
of interesting correlations. }

The term in \eqref{e:eeeee}, though, is always positive when $h/2$ is in the support of
$\kappa_\sigma$. 
We can avoid this term by using the following `leave-out' estimator
\begin{equation}
\bar \gamma(h)= \int_{W \cap W_{-h}} \sum_{u,v \in X\cap W}^{\neq} 
 \frac{\kappa_\sigma(z-u) \kappa_\sigma(z + h - v)}{w(z;u) w(z+h;v)} \dd z, \label{e:gammabar}
\end{equation}
where leave-out refers to omitting `diagonal terms'  $u=v$  in
$\hat \rho(z) \hat \rho(z+h)$ (with $u,v
\in X\cap W$).
Similarly, when $X$ is isotropic, an estimator of $\gamma^\mathrm{iso}$ can be defined in terms of $\bar \gamma$, as
\begin{equation}
\bar \gamma^\mathrm{iso} (r)= r^{d-1}\int_{S^{d-1}} \bar \gamma (rs) \dd \nu_{d-1}(s) . 
\label{e:gammabariso}
\end{equation}

For the global $K$-function estimators, $\gamma$ is evaluated at $y-x$
for distinct $x, y\in X\cap W$. In this case the relevant
expectation is $\mathrm{E}\{\bar\gamma(y-x) \mid x,y\in X\}$. As in Section~\ref{s:local-bias} we
obtain this by considering the second-order reduced Palm distribution
at distinct $x,y\in W$ with $\rho^{(2)}(x,y)>0$, by assuming that $X$ has $n$-th
order intensity functions $\rho^{(n)}$ for $n\le 4$, and by neglecting
the edge corrections for simplicity: 
\begin{align*}
\text{E}\{ \bar\gamma(y-x) & \mid  x,y\in X \} =
\\ \int_{W \cap W_{-(y-x)}}
& \bigg(\int_W\int_W \frac{\rho^{(4)}(x,y,u,v)}{\rho^{(2)}(x,y)} \kappa_\sigma(z-u)\kappa_\sigma(z + (y-x) - v) \dd u \dd v 
\\&+ \kappa_\sigma(z-x)^2 + \kappa_\sigma(z-y)\kappa_\sigma(z + y-2x)
\\ &+ \int_W\frac{\rho^{(3)}(x,y,u)}{\rho^{(2)}(x,y)} \big[
\{\kappa_\sigma(z-x)+\kappa_\sigma(z -y)\}\kappa_\sigma(z + (y-x) - u)
\\ &\quad +\kappa_\sigma(z-u)\{\kappa(z -x) + \kappa_\sigma(z + y-2x)\}\big] \dd u \bigg)
\dd z.
\end{align*}
{Again, in case of a Poisson process,
$\rho^{(4)}(x,y,u,v)/\rho^{(2)}(x,y)=\rho(u)\rho(v)$ and the first term is approximately $\gamma(y-x)$, subject to the subtleties discussed above.}
The other three terms are related to the terms with $u,v\in\{x,y\}$ of the double sum in 
\eqref{e:gammabar}, and yield a positive bias. We expect this bias to be small when
$\sigma$ is reasonably small, since the excess terms become negligible
far from $x$ and $y$, and the integral is over all of $W \cap W_{-h}$.
The three terms could be avoided by considering the
  further modified `leave-one pair-out' estimator
\[ \tilde \gm(h;x,y)= \int_{W \cap W_{-h}} \sum_{u,v \in (X\cap W)\setminus \{x,y\}}^{\neq} 
 \frac{\kappa(z-u) \kappa(z + h - v)}{w(z;u) w(z+h;v)} \dd z,\qquad\mbox{ with $h=y-x$},
\]
but this depends on $(x,y)$ not only through $h=y-x$ which precludes the
use of interpolation schemes as discussed in Section~\ref{sec:computation}.

In case of two point processes we just use
\[ \hat \gm_{12}(h)= \int_{W \cap W_{-h}} \hat \rho_1(z) \hat \rho_2(z
  + h)\,\mathrm d z\]
for kernel estimators $\hat \rho_1$ and $\hat \rho_2$, since in this
case almost surely there are no diagonal terms $u=v$ in $\hat \rho_1(z) \hat \rho_2(z + h)$ (with $u \in
X_1$ and $v \in X_2$).

\subsection{Computation of $\gamma$ and $\gamma^\text{iso}$}\label{sec:computation}

We compute $\gamma(h)$ for a given intensity function $\rho$ using a simple Monte Carlo
integration algorithm: we generate uniform random samples $U_i$, $i=1,\ldots,n$,
on $W \cap W_{-h}$ and approximate $\gamma(h)$ by the
unbiased Monte Carlo estimate
\begin{equation}\label{e:mmmmm} \gm_{MC}(h)=\frac{|W \cap W_{-h}|}{n}
\sum_{i=1}^n 
\rho(U_i)\rho(U_i + h). 
\end{equation}
To achieve a desired precision, we consider the standard error
  $\sigma_\text{MC}/\sqrt{n}$ of $\gm_{MC}(h)$ and choose $n$
  so that the coefficient of variation becomes less than a selected
  threshold $\alpha$: $\sigma_{MC}/(\sqrt n \mu_\text{MC})<\alpha$. 
  For the simulation studies in Section~\ref{s:sim}, we used $\alpha=.001$ or $\alpha=.005$.
  In practice, we wish to evaluate $\gamma$ at many values of $h$. Thus it is
  convenient to generate a single sequence of random samples $V_j, j=1,\ldots,n'$ on $W$,
  and for each $h$ use a subsequence $\{U_i^{(h)}\} = \{V_j \mid V_j \in W \cap W_{-h}\}$.
  We choose $n'$ sufficiently large to produce the requisite length of sub-sequence for each
  $h$.

For $\gamma^\text{iso}(r)$, we follow a similar approach, generating also
random independent $s_i$ uniformly on $\{s \mid s \in \mathbb{S}^{d-1}, U_i + rs \in W\}$, and computing
\begin{equation}\label{e:nnnnn}
\gamma_\text{MC}^\text{iso} = \frac{\int_{\mathbb{S}^{d-1}} |W \cap W_{-rs}| \dd \nu_{d-1}(s)}{\varsigma_d n}
\sum_{i=1}^{n}
\rho(U_i)\rho(U_i + r s_i). 
\end{equation}
The integral $\int_{\mathbb{S}^{d-1}} |W \cap W_{-rs}| \dd \nu_{d-1}(s)$ is easy to compute
when $W$ is a rectangular window. As above, $U_i$ and $s_i$ are typically generated
for each $r$ as appropriate subsequences of shared larger sequences $V_j$ and $t_j$, respectively,
sampled uniformly on $W$ and $\mathbb{S}^{d-1}$, respectively.

In practice $\rho$ is replaced by an estimate. Then for the kernel-based leave-out estimator \eqref{e:gammabar}, 
$\rho(U_i)\rho(U_i + h)$ in \eqref{e:mmmmm} is replaced by 
\[\sum_{u,v \in X\cap W}^{\neq}  \frac{\kappa_\sigma(U_i-u) \kappa_\sigma(U_i + h - v)}{w(z;u) w(z+h;v)},\] 
which is evaluated using a fast routine written in C. In a similar way, when
$X$ is isotropic and \eqref{e:gammabariso} is used, $\rho(U_i)\rho(U_i + rs_i)$
in \eqref{e:nnnnn} is replaced by a double sum.

Since $\gamma$ and $\gamma^\text{iso}$ are quite smooth, it is possible
to interpolate them very accurately based on a moderate number of points $h_j$
or $r_j$. This is especially helpful for $\gamma^\text{iso}$ because it is
one-dimensional. For the kernel-estimated $\bar\gamma^\text{iso}$ or
$\hat\gamma^\text{iso}$, we find that linear interpolation based on sample
spacing of $|r_{j+1} - r_j| < \sigma/10$ gives estimates within .01\% of the
true values.
The interpolation scheme is especially helpful for the $K$-functions as the number of points
grows large, in which case we must evaluate $\gamma$ (or $\gamma^\text{iso}$ in the isotropic case) at a very
  large number of pairs of points.

The proposed Monte Carlo computation becomes very slow when
especially precise coefficient of variation $\alpha$ is desired, or when using
kernel-based estimates with very small kernel bandwidth $\sigma$ or large
number of points $N$. For these cases, it may be beneficial to apply a
variance reduction technique such as antithetic variables, or to
consider an
approximate convolution based on discrete Fourier transforms, with a
kernel-based estimate of $\rho$, when desired, based on quadrat counts. When
the side length of the quadrats is much less than $\sigma$, we expect this
method to produce
accurate estimates of $\gamma$ (or $\gamma^\text{iso}$ in the isotropic case).

\section{Simulation study}\label{s:sim}

To compare global and local estimators for $K$ and $g$, we simulated 100 point
patterns on the unit square $W=[0,1]^2$ for each of nine point process models obtained by combining
three different types of point process interactions with four types
of intensity functions. For plots of estimated $K$ or $g$ we simulated
a further 1000 point patterns of the considered point process model. 

More specifically we simulated stationary point processes of the types
 Poisson (no interaction), log-Gaussian Cox
\cite[LGCP -- these are clustered/aggregated, see][]{moeller:syversveen:waagepetersen:98}, and
determinantal \cite[DPP -- these are regular/repulsive, see][]{lavancier:moeller:rubak:12},
and subsequently subjected them to independent thinning
to obtain various types of intensity functions. Note that independent
thinnings of stationary point processes are soirs
\cite[cf.][]{BaddeleyNon2000}. The intensities of the stationary point
processes were adjusted to obtain on average 200 or 400 points in the
simulated point patterns (that is, after independent thinning). 

For the Gaussian random field underlying the LGCP we used an exponential covariance function with unit variance and correlation scale $0.05$ resulting in the isotropic pair correlation function
\[g_\mathrm{LGCP}(r) = \exp\{\exp(-r/.05)\} .\]
For the DPP we used a Gaussian kernel with scaling parameter $\alpha=0.02$ leading to
\[g_\mathrm{DPP}(r) = 1 - \exp\left\{-2(r/.02)^2\right\}.\]

\begin{figure}
\centering
\includegraphics[width=5in]{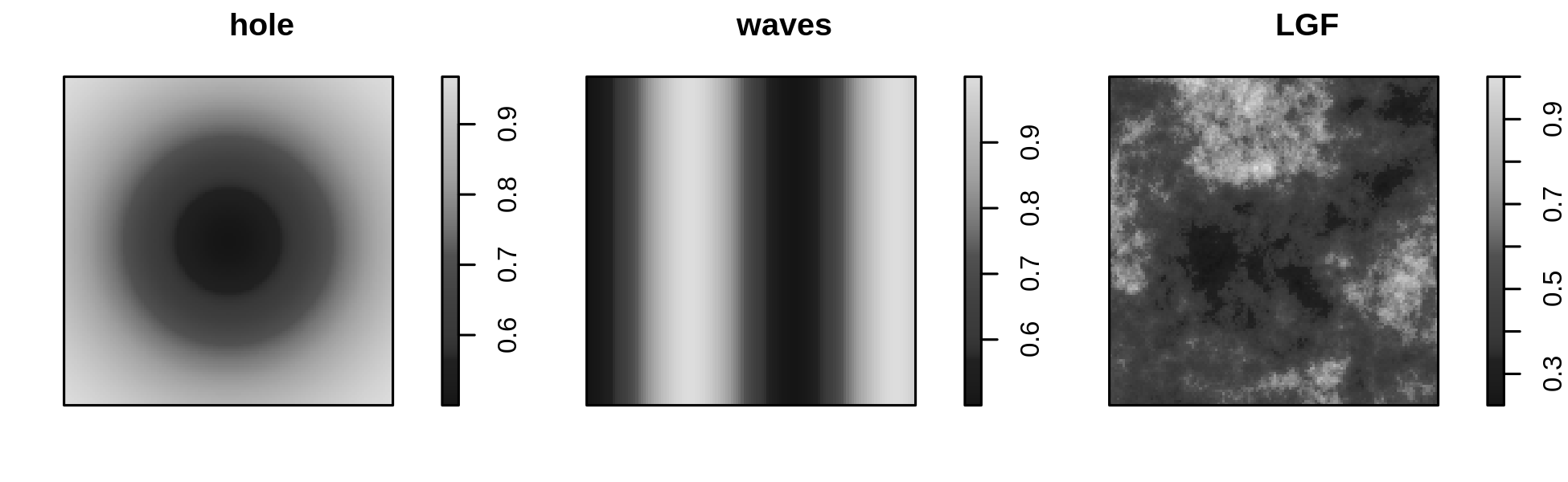}
\caption{Plots of the `hole', `waves' and `LGF' thinning profiles.}
\label{f:profs}
\end{figure}

The intensity functions were of type constant (no thinning), `hole', `waves', or log-Gaussian random field (`LGF').  Intensity
functions of the `hole' and `waves' types were obtained by independent thinning using
spatially varying retention probabilities 
\begin{align*}
p_\mathrm{hole}(x,y) &= 1 - .5 \exp\left[ - \left\{(x - .5)^2 + (y - .5)^2\right\}/.18\right], \\
  p_\mathrm{waves}(x,y) &= 1 - .5 \cos^2 (5x),\\
p_\mathrm{LGF}(x,y) &= \lambda(x,y)/\sup_{(u,v) \in W} \lambda(u,v),
\end{align*}
for $(x,y)\in [0,1]^2$. In case of `LGF', $\log \lambda$ was generated as a realization of a Gaussian random field with exponential
  covariance function, with variance .1 and correlation scale .3. The resulting `LGF' retention probability surface is much less smooth than for `hole' and `waves' but similar to `hole' and `waves' in terms of intensity contrast and spatial separation of high-intensity and low-intensity
regions.
The surfaces of retention probabilities are shown in Figure \ref{f:profs}.

Simulations were carried out and analyzed using the \texttt{R} package
\texttt{spatstat}, and a new package \texttt{globalKinhom} that implements the
global $K$- and pair correlation function estimators using Monte-Carlo
estimates of $\gamma$ as described in Section \ref{sec:computation}
\citep{Rlang,baddeley:rubak:turner:15,globalKinhom}.
In most cases we set the precision of the Monte-Carlo estimates to $\alpha = .005$. When
probability intervals and root integrated mean square error (RIMSE) values are
shown, we use $\alpha = .001$ instead, where the more precise calculation
produced slightly smaller RIMSE values. We also tested smaller values of
$\alpha$ in a few particular cases, and did not observe any reduction in RIMSE
values below $\alpha = .001$.
We do not show simulation results for all scenarios
since in many cases the different scenarios led to qualitatively similar
conclusions.

To investigate our cross $K$ and cross pair correlation function estimators we generated simulations from a bivariate LGCP detailed in Section~\ref{sec:crosssim}.

\subsection{Estimation of $K$ and pair correlation functions}

We initially compare the bias of global and local estimators of
the $K$-function using in both cases kernel estimators of the
intensity function obtained with a Gaussian kernel with bandwidth
$\sigma$ chosen by the method of Cronie \& van Lieshout (2018)\nocite{CronieNonmodelbased2018}, as
implemented in the \texttt{spatstat} procedure \texttt{bw.CvL} (CVL
for convenience in the following). The selected bandwidths vary
around .05 (see third column in Table~\ref{t:bws-one}), with slightly larger
bandwidths for LGCP than for Poisson
and DPP. For the global estimator we consider the isotropic estimator
\eqref{e:isoKglobal}, since the pair correlation functions of the
point processes tested here are all isotropic, as in the setting of
Section \ref{s:isoK}, and the estimation of $\gamma^\mathrm{iso}$ is
less computationally intensive than that of $\gamma$. We consider both
the estimator \eqref{e:kernelgamma} and the leave-out estimator
\eqref{e:gammabar} of the function $\gamma$. Similarly we also
consider the local estimator using either the original kernel
estimator \eqref{e:kernelrho} or the leave-out estimator
\eqref{e:leaveoutkernel} suggested in \cite{BaddeleyNon2000}. 

For better visualization of the simulation results we transform the $K$-function estimators into estimators of the so-called $\{L(r)-r\}$-function via the one-to-one transformation
\[L(r)-r = \sqrt{K(r)/\pi}-r.\]
We only show results in case of the waves intensity function with on average 400 simulated points, since the results for the other intensity functions and with on average 200 simulated points give the same qualitative picture.
\begin{figure}
  \centering
  \begin{tabular}{ccc}
    \includegraphics[width=0.33\textwidth]{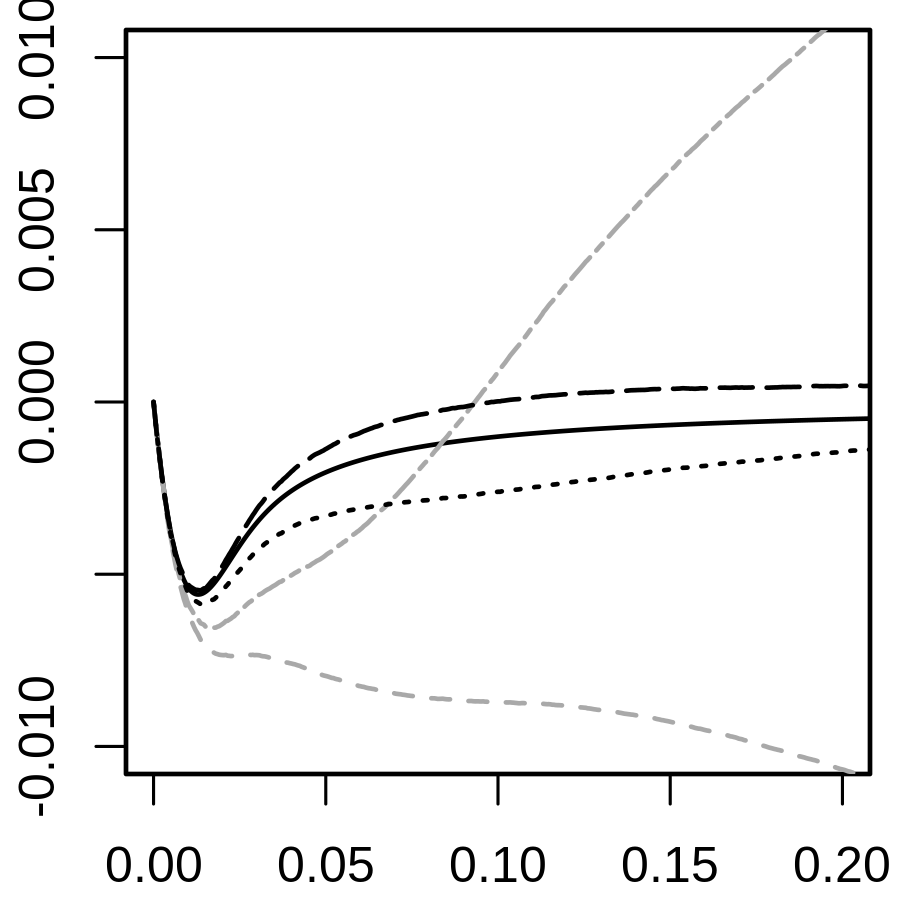}
    \includegraphics[width=0.33\textwidth]{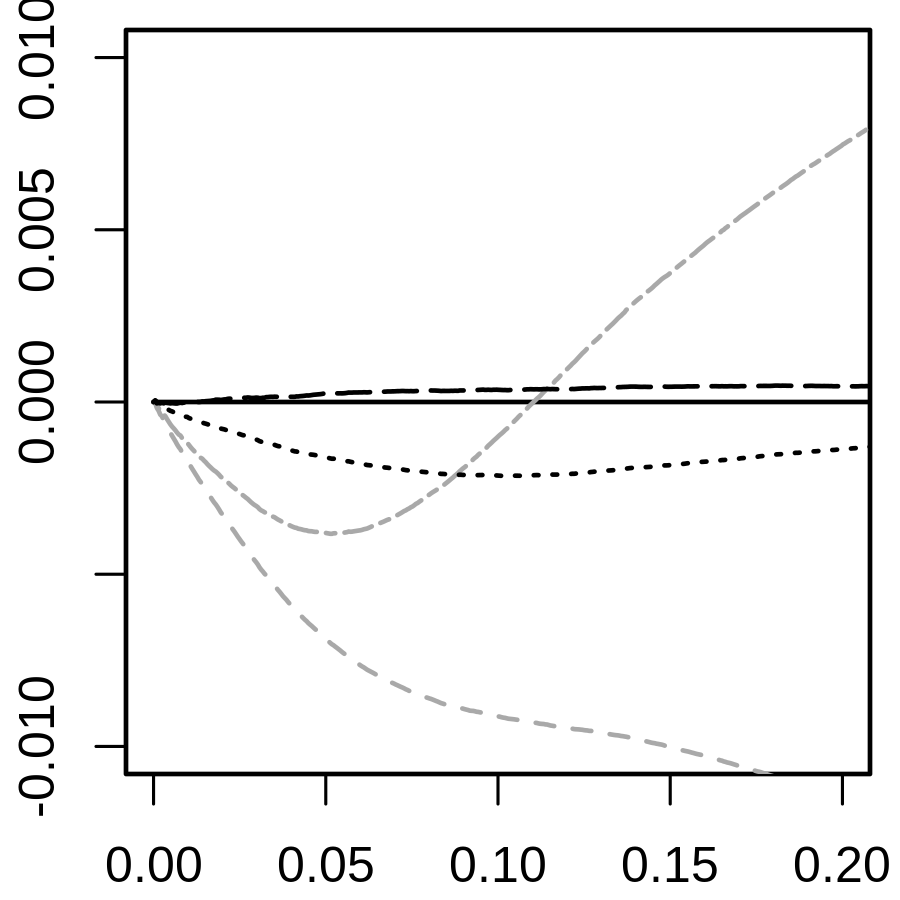}
    \includegraphics[width=0.33\textwidth]{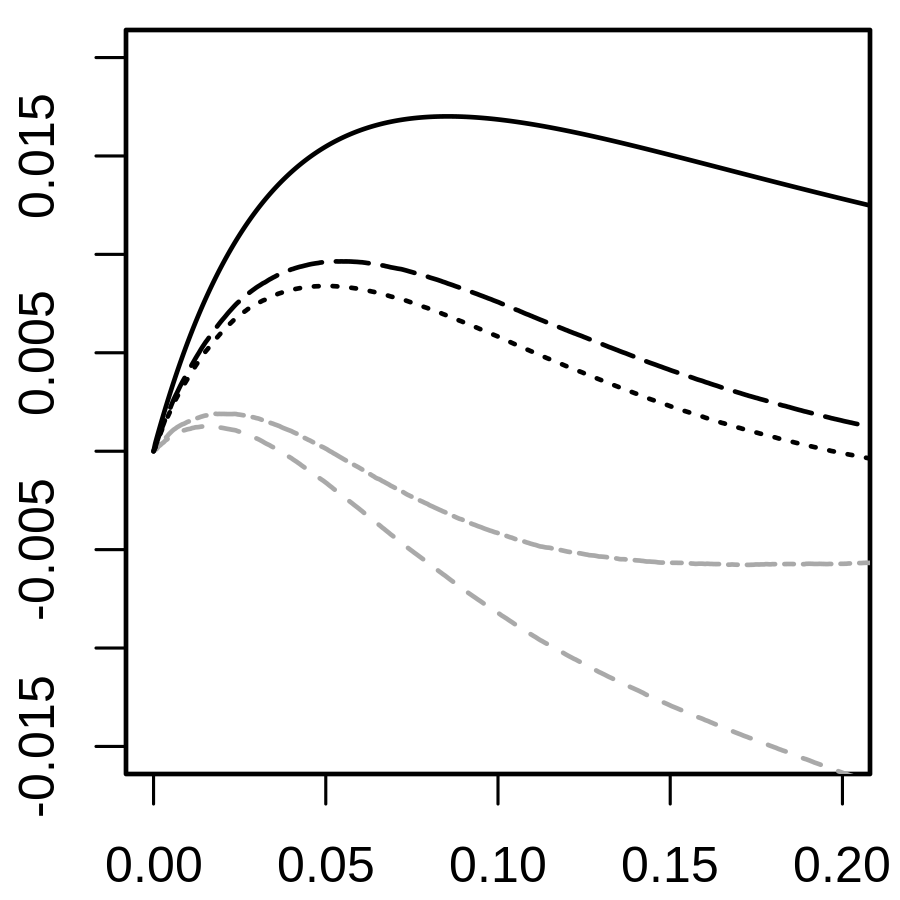}
    \end{tabular}
    \caption{Averages of estimates of $L(r)-r$ obtained from simulations in case of the waves intensity function with 400 simulated points on average. Left to right: DPP, Poisson, LGCP. The estimates are obtained using $\hat K_\mathrm{global}^\mathrm{iso}$ with or without the leave-out approach
    (\protect\includegraphics[width=1.2cm]{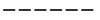}, 
    \protect\includegraphics[width=1.2cm]{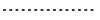}, respectively)
    or $\hat K_\mathrm{local}$ with or without the leave-out approach
    (\protect\includegraphics[width=1.2cm]{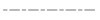},
    \protect\includegraphics[width=1.2cm]{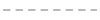}, respectively)
    for kernel estimation of $\gamma$ or the intensity function.
    True values of $L(r) - r$ are shown for comparison
    (\protect\includegraphics[width=1.2cm]{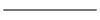}).}
    \label{f:Lr-r}
  \end{figure}

  Figure~\ref{f:Lr-r} shows averages of the simulated estimates and it is
obvious that the global estimators are much less biased than the local
estimators.  It is clearly advantageous to use the leave-out versions
for the global estimator. The leave-out approach is also advantageous for
the local estimator, at least for small distances $r$.
The biases of the leave-out local estimator are as discussed in
Section~\ref{s:local-bias}: strong negative bias at short distances due to the
covariance of $\bar \rho(x)$ and $\bar \rho(y)$, and
strong positive bias at large distances due to Jensen's inequality $\text{E}(1/\bar\rho(x))
>1/\text{E}(\bar \rho(x))$.
The leave-out global estimator appears to be close to unbiased in case of DPP and Poisson
but is too small on average in case of LGCP.

\begin{table}
\caption{ Mean ($\pm$ st. dev.) of CVL and LCV bandwidths, for each type of spatial
point process we considered. The expected number
of points for each listed process is 400. } \label{t:bws-one}
    \centering
    \begin{tabular}{ccccc}
Interaction type & Intensity function
&  $\sigma_\text{CVL}$ &  $\sigma_\text{LCV}$ \\
\hline
DPP  &  constant & 0.046 (0.005) & 0.63 (0.15) \\
  &  hole  & 0.045 (0.004) & 0.33 (0.22) \\
  & waves & 0.048 (0.004) & 0.28 (0.25) \\
  & LGF & 0.047 (0.005) & 0.22 (0.16) \\
Poisson  &  constant & 0.047 (0.006) & 0.59 (0.21) \\
  &  hole  & 0.048 (0.007) & 0.29 (0.23) \\
  & waves  & 0.050 (0.006) & 0.14 (0.11) \\
  &  LGF & 0.050 (0.006) &  0.17 (0.13) \\
LGCP  &  constant & 0.066 (0.009) & 0.040 (0.007) \\
 &   hole & 0.064 (0.012) & 0.044 (0.008) \\
 &  waves & 0.071 (0.011) & 0.042 (0.008) \\
 &  LGF  &  0.066 (0.011) & 0.042 (0.007) \\
    \end{tabular}
\end{table}

There exist a number of alternatives to the CVL approach to choosing the bandwidth for the kernel estimation. We therefore also investigate bias in the
case where the bandwidth is selected using the likelihood cross validation
(LCV) method implemented in the \texttt{spatstat} procedure
\texttt{bw.ppl}. Results regarding the LCV selected bandwidths are
summarized in the fourth column of
Table~\ref{t:bws-one}. Comparison of
the CVL and LCV results in Table~\ref{t:bws-one} shows
that the LCV approach tends to select considerably larger bandwidths $\sigma$ than
the CVL method for the DPP and Poisson process, and somewhat smaller $\sigma$ for the
LGCP.

\begin{figure}
  \centering
  \begin{tabular}{ccc}
    \includegraphics[width=0.33\textwidth]{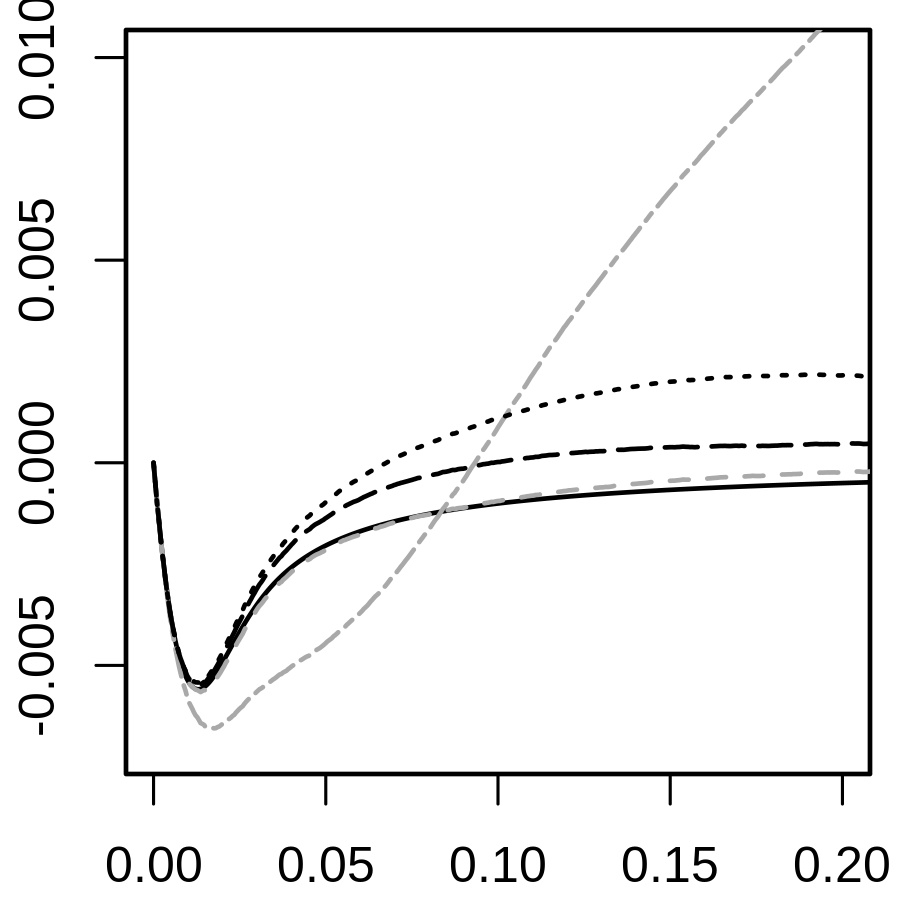}
    \includegraphics[width=0.33\textwidth]{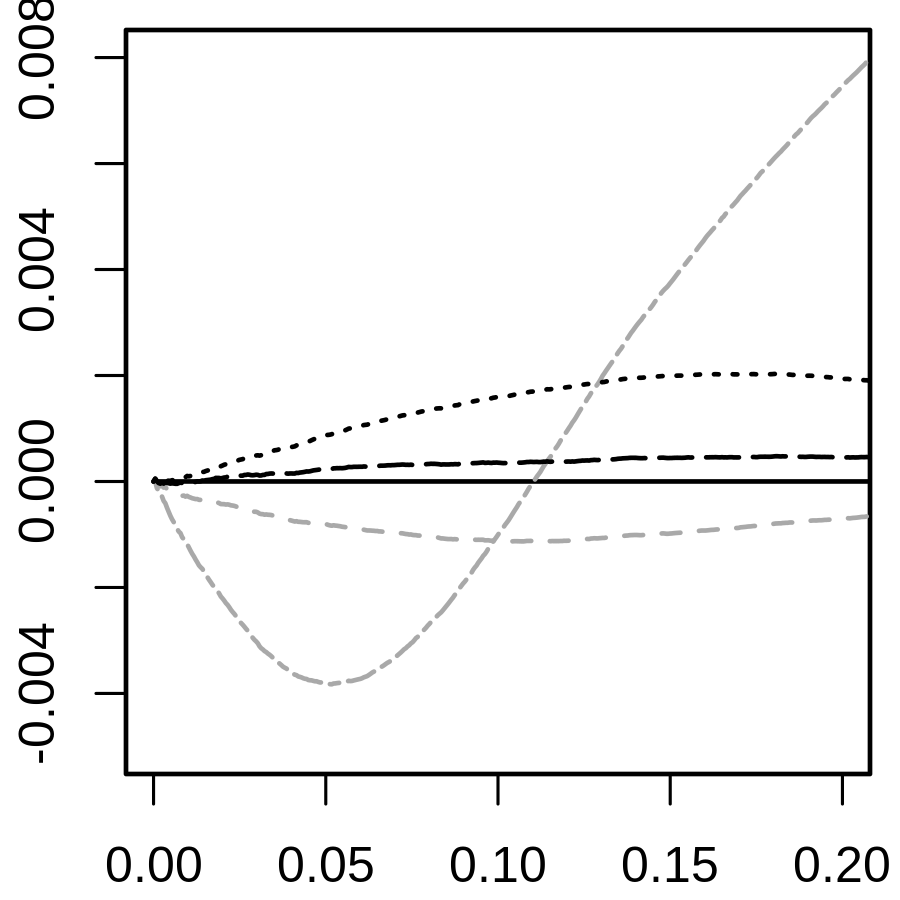}
    \includegraphics[width=0.33\textwidth]{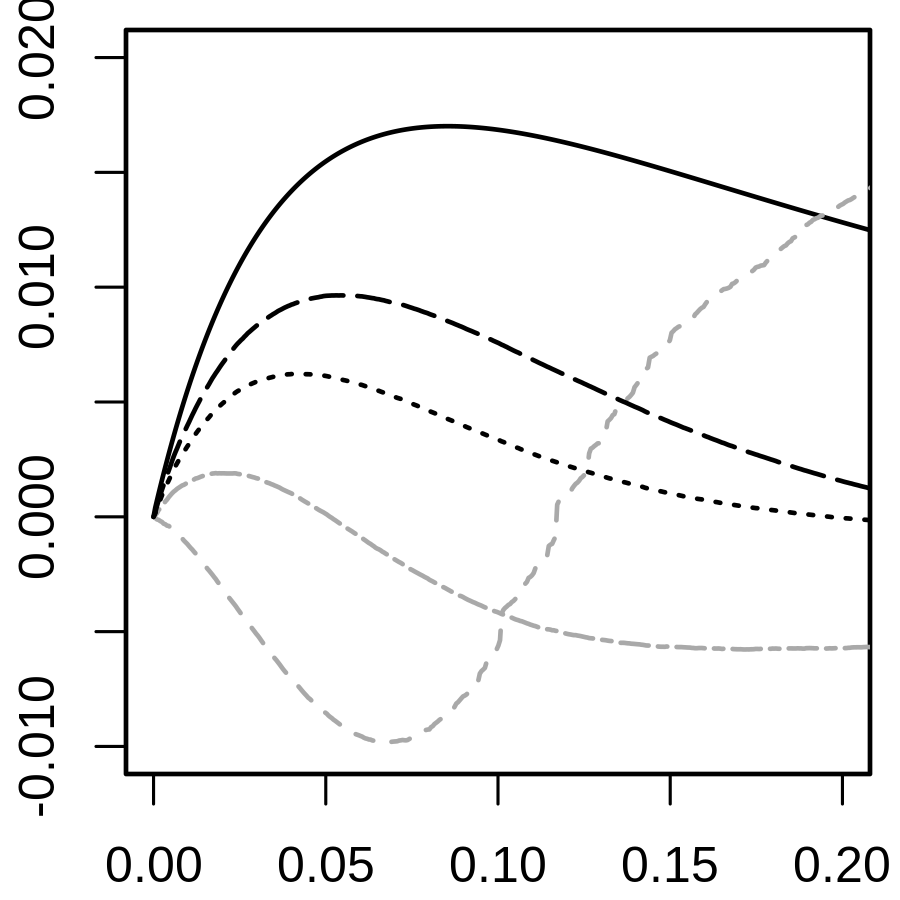}
    \end{tabular}
    \caption{Averages of estimates of $L(r)-r$ obtained from simulations in case of the waves intensity function with 400 simulated points on average. Left to right: DPP, Poisson, LGCP. The estimates are obtained using the global
    (\protect\includegraphics[width=1.2cm]{caption/t5w1-5b.png}~CVL,
    \protect \includegraphics[width=1.2cm]{caption/t3w1-5b.png}~LCV)
    or local
    (\protect\includegraphics[width=1.2cm]{caption/t6w1-5g.png}~CVL,
    \protect \includegraphics[width=1.2cm]{caption/t2w1-5g.png}~LCV)
    estimators of the $K$-function with either CVL or LCV for selecting the
    bandwidth (in all cases the leave-out approach is
    used).
    True values of $L(r) - r$ are shown for comparison
    (\protect\includegraphics[width=1.2cm]{caption/t1w1b.png}).}
    \label{f:cvlvsppl}
  \end{figure}

Figure~\ref{f:cvlvsppl} compares averages of the global and local estimators
using either of the two approaches to bandwidth selection and with
leave-out in all cases.
Again we show only results for the
waves intensity function and expected number of points equal to 400.
The bias of the estimators is quite
sensitive to the choice of bandwidth selection method. In case of DPP and Poisson, the global estimator using CVL and the local
estimator using LCV perform similarly with the global estimator a bit more biased than the local for DPP and vice versa for Poisson. The global estimator
performs slightly worse when combined with LCV than with CVL, likely due to
the inherent biases of the kernel estimator $\bar \rho$, which become
more pronounced as $\sigma$ increases.
The local estimator with CVL is strongly biased for almost all $r$ considered.
The improved performance with LCV is likely due to the reduced variances and
covariances for $\bar \rho$ when a larger bandwidth is used. This also
explains the strong bias of the local estimator with LCV for the LGCP, since
$\sigma_\mathrm{LCV}$ is typically smaller than $\sigma_\mathrm{CVL}$ in
that case.
The global estimator for the LGCP has the smallest bias with the CVL
method and has  much less bias than the local
estimator regardless of whether CVL or LCV is used. It is not surprising
that the LGCP is the most challenging case for both the global and local
estimators, since the random aggregation of the LGCP tends to be entangled
with the variation in the intensity function.

\begin{figure}
  \centering
  \begin{tabular}{ccc}
    \includegraphics[width=0.33\textwidth]{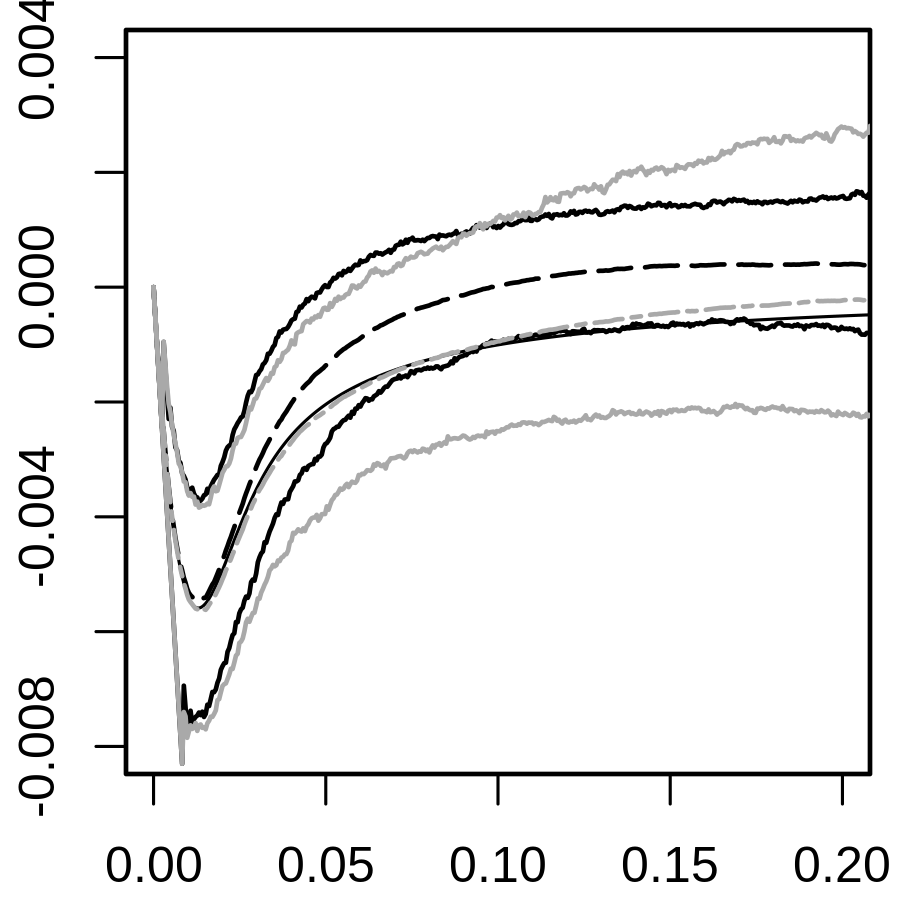}
    \includegraphics[width=0.33\textwidth]{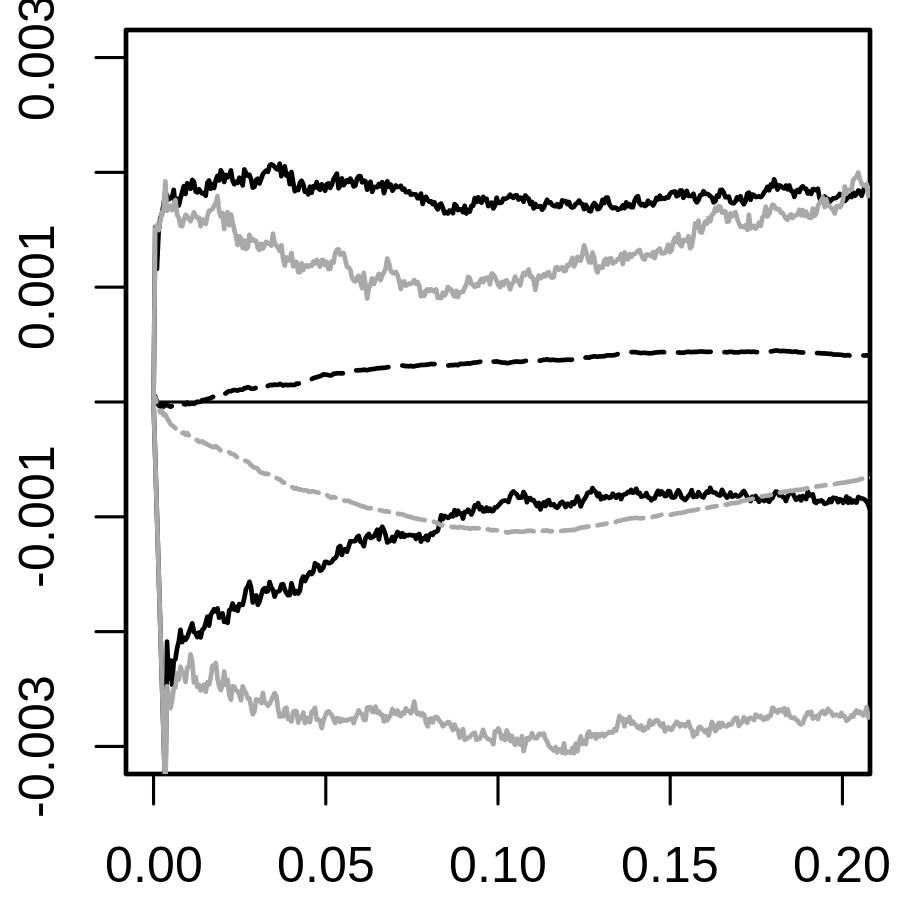}
    \includegraphics[width=0.33\textwidth]{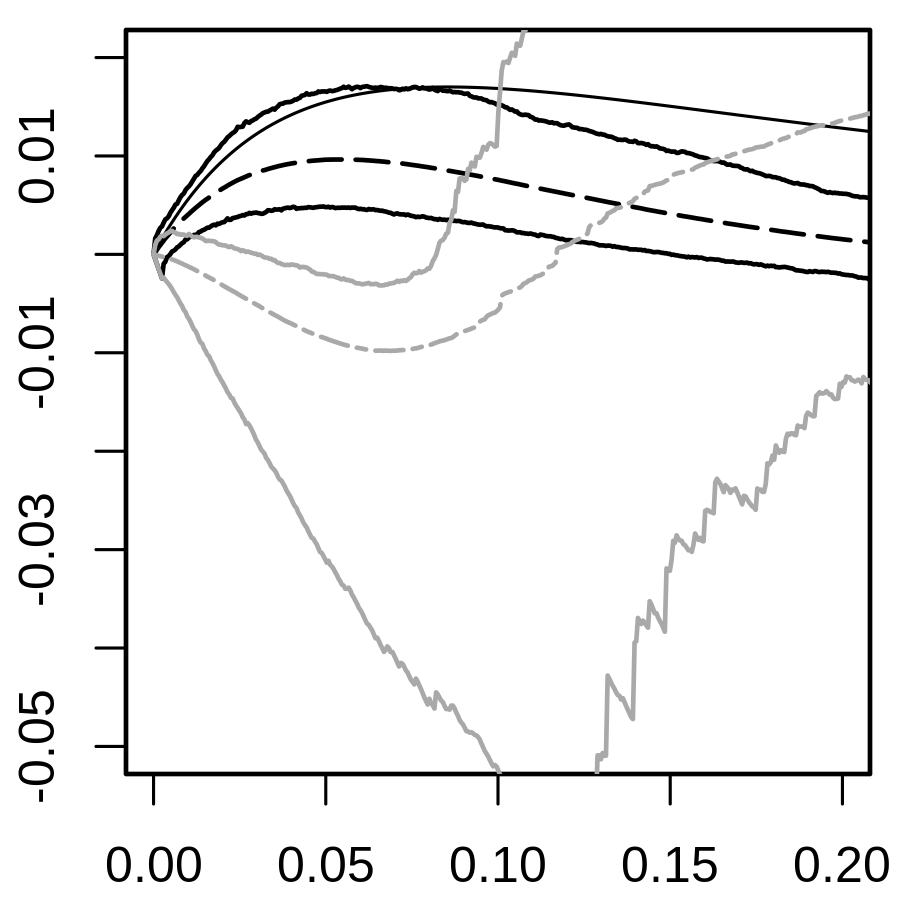}
    \end{tabular}
    \caption{Averages and 95\% pointwise probability intervals for estimates of $L(r)-r$
    in case of the waves intensity function with 400 simulated points on
    average. Left to right: DPP, Poisson, LGCP. The estimators used are the
    leave-out global estimator using CVL
    (\protect\includegraphics[width=1.2cm]{caption/t5w1-5b.png})
    and the leave-out local estimator using LCV.
    (\protect\includegraphics[width=1.2cm]{caption/t6w1-5g.png}), with pointwise probability
    intervals shown in like shade. True values of $L(r) - r$ are also shown
    (\protect\includegraphics[width=1.2cm]{caption/t1w1b.png}).}
 \label{f:var}
\end{figure}

We finally compare the sampling variability of the leave-out global
estimator using CVL and the leave-out local estimator using LCV.
Figure~\ref{f:var} shows 95\% pointwise probability intervals and averages for
the two estimators, again with 400 simulated points on average and the `waves'
intensity function, and Table~\ref{t:imse-one} gives root integrated mean
square error (RIMSE) values for the $K$-function estimators applied to
each process, for each combination of CVL or LCV with the local or global
leave-out estimator. Figure~\ref{f:var} indicates that the global estimator has smaller variance than the local estimator. This should also result in smaller mean square error for Poisson and LGCP where the bias is also smallest for the global estimator. For DPP the picture is not completely clear regarding mean square error since in this case the global estimator has larger bias than the local estimator. 
Table~\ref{t:imse-one} gives more insight where a first observation is
that the leave-out local estimator is very sensitive to the choice of bandwidth
selection method with LCV performing much better than CVL for DPP and Poisson
and vice versa for LGCP. The leave-out global estimator is much less sensitive
to choice of bandwidth selection method. Best results in terms of RIMSE are
obtained with the leave-out global estimator combined with CVL.

\begin{table}

\caption{RIMSE $\times 10^2$ of local and global
$K$-function estimators with CVL and LCV bandwidths.}\label{t:imse-one}

\centering

\begin{tabular}{cccccc}
&&\multicolumn{2}{c}{$\hat K_\text{local}$} & \multicolumn{2}{c}{$\hat K_\text{global}$} \\
Interaction type & Intensity function &  CVL & LCV & CVL & LCV \\
\hline
    DPP &  flat & 0.59 & 0.069 & 0.029 & 0.060  \\
        &  hole & 0.64 & 0.107 & 0.031 & 0.128  \\
        & waves & 0.60 & 0.052 & 0.049 & 0.121 \\
        &  LGF  & 0.59 & 0.060 & 0.050 & 0.110 \\
Poisson &  flat & 0.45 & 0.083 & 0.028 & 0.069  \\
        &  hole & 0.45 & 0.120 & 0.034 & 0.103  \\
        & waves & 0.40 & 0.061 & 0.037 & 0.093 \\
        &  LGF  & 0.37 & 0.087 & 0.050 & 0.089 \\
   LGCP &  flat & 0.89 & 0.999 & 0.573 & 0.628  \\
        &  hole & 0.87 & 1.554 & 0.576 & 0.636  \\
        & waves & 0.89 & 1.146 & 0.528 & 0.613 \\
        & LGF   & 0.90 & 1.506 & 0.542 & 0.625 \\
\end{tabular}

\end{table}

\begin{figure}
  \centering
  \begin{tabular}{ccc}
    \includegraphics[width=0.33\textwidth]{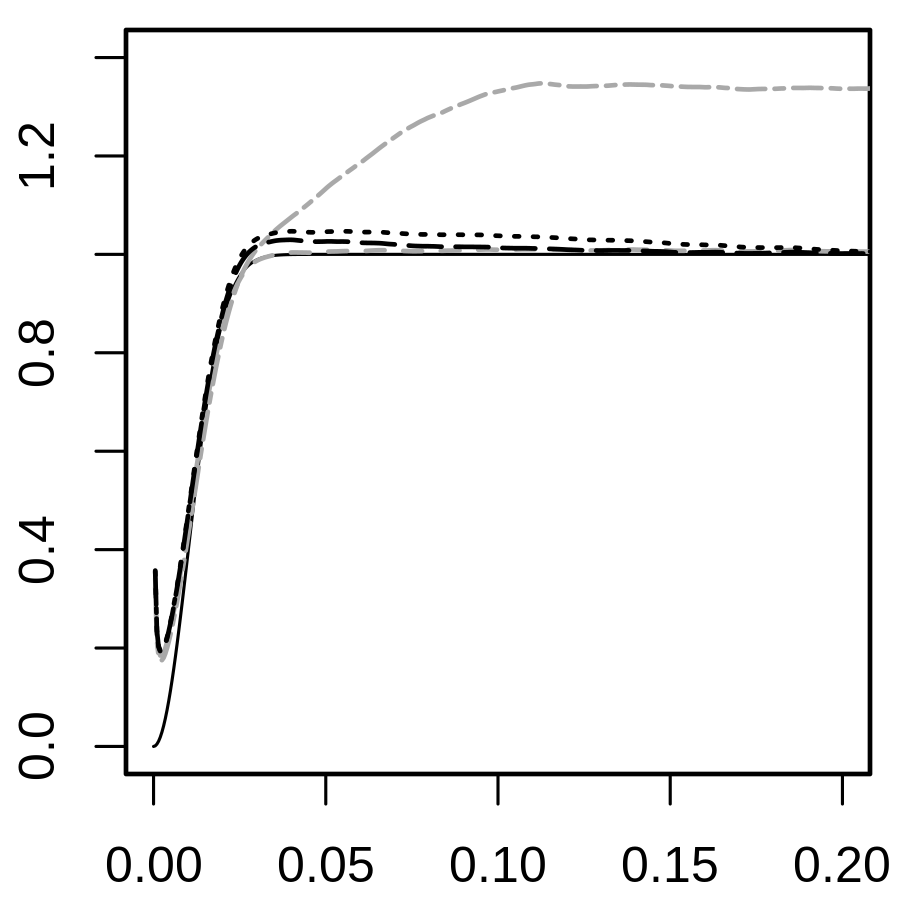}
    \includegraphics[width=0.33\textwidth]{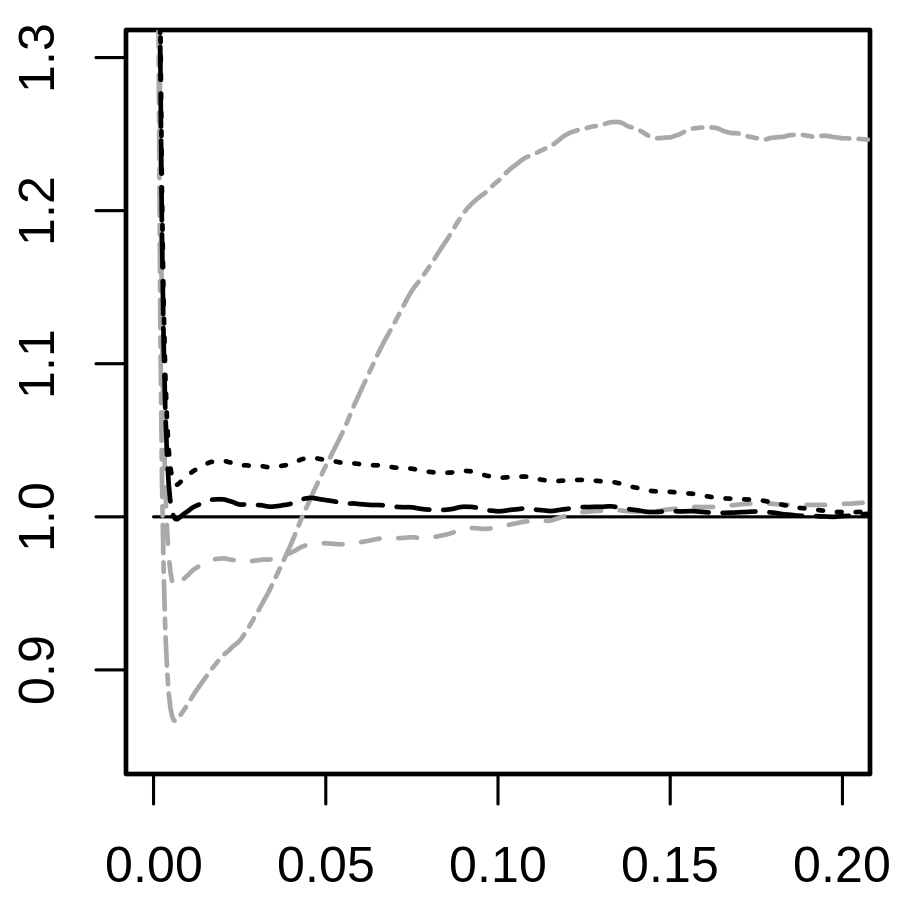}
    \includegraphics[width=0.33\textwidth]{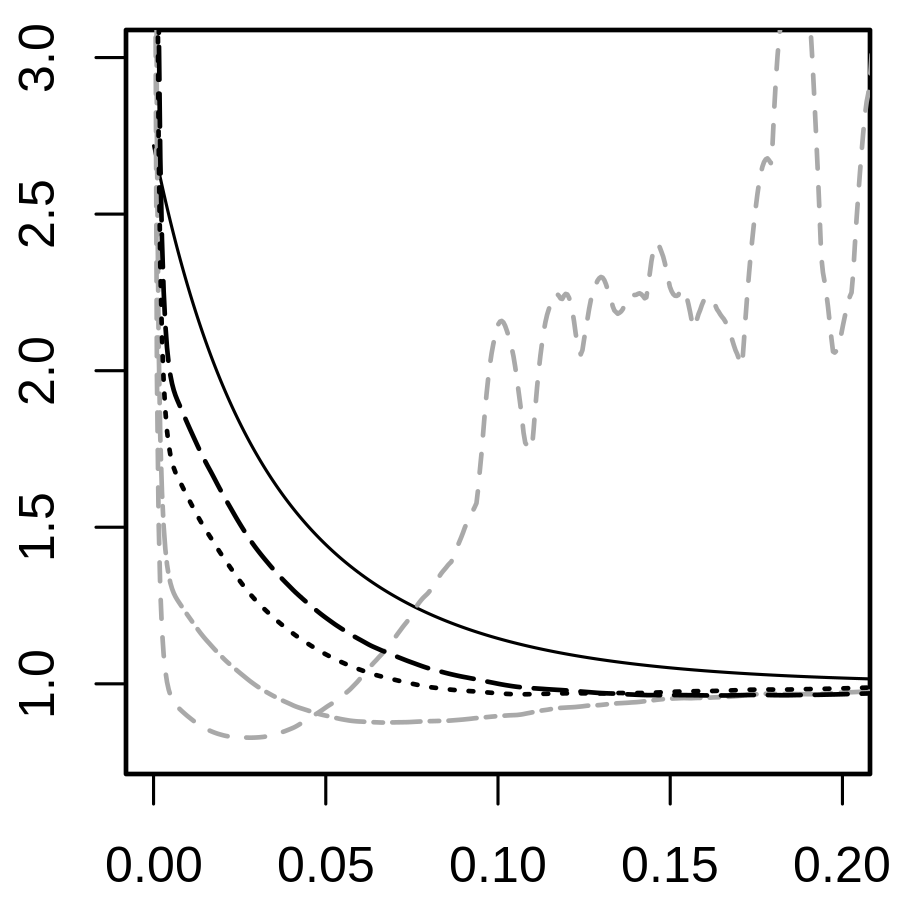}
    \end{tabular}
    \caption{Averages of estimates of $g_1(r)$ obtained from simulations in case of the waves intensity function with 400 simulated points on average. Left to right: DPP, Poisson, LGCP. The estimates are obtained using the global
    (\protect\includegraphics[width=1.2cm]{caption/t5w1-5b.png}~CVL,
    \protect \includegraphics[width=1.2cm]{caption/t3w1-5b.png}~LCV)
    or local
    (\protect\includegraphics[width=1.2cm]{caption/t6w1-5g.png}~CVL,
    \protect \includegraphics[width=1.2cm]{caption/t2w1-5g.png}~LCV)
    estimators of the pair correlation function with either CVL or LCV bandwidth selection. (In each case, the
    leave-out approach is used.)
    True values of $g(r)$ are shown for comparison
    (\protect\includegraphics[width=1.2cm]{caption/t1w1b.png}).}
    \label{fig:ginhom}
  \end{figure}

  Figure~\ref{fig:ginhom} shows averages of leave-out global and local
estimators of the isotropic pair correlation function using either CVL or LCV in case of the wave intensity
with 400 points on average. Once again, local estimators are most strongly
biased with the bandwidth selection method that produces the smaller
bandwidth: CVL for the DPP and Poisson processes, and LCV for the LGCP.
The bias is small to moderate for the global estimators with largest
bias in case of LGCP. For the DPP and Poisson case positive
bias of the local and global
estimator occurs for very small distances.

  \subsection{Estimation of cross $K$ and cross pair correlation functions}\label{sec:crosssim}

  To investigate the cross  $K$ and cross pair correlation function estimators, we simulated 100
bivariate point patterns for each model of a bivariate point process $(X_1,X_2)$, where either $X_1$ and $X_2$ are independent or
display segregation or co-clustering. Processes that were
chosen for plotting were simulated an additional 1000 times. Inhomogeneous
intensity functions were subsequently obtained using independent thinning of stationary bivariate point processes, 
where the two point processes have the same intensity, and 
 the constant, `hole', and `waves' retention probabilities $p$ as described in connection to Figure~\ref{f:profs} were used.
 This implies $\rho_1(x) = \rho_2(x)$ for $x\in[0,1]^2$ (we did not investigate
any scenarios where $\rho_1 \neq \rho_2$). 

In the case of independence, $X_1$ and $X_2$ are independent Poisson
processes. For the dependent cases, we considered a bivariate LGCP. Specifically, for $i=1,2$, $X_i$ has random intensity function 
\[ \Lambda_i(u)=p(u)\exp\{\mu_i+ \alpha_i Y(u) + \beta U_i(u)\}, \qquad i=1,2,\]
where $Y$, $U_1$, and $U_2$ are independent zero-mean unit-variance Gaussian
random fields with isotropic exponential correlation functions given by $\exp(-r/\phi)$ and
$\exp(-r/\psi_i)$ ($r\ge0$), $i=1,2$, respectively, and where $\mu_i\in\mathbb R$, $\alpha_i\in\mathbb R$, and $\beta>0$ are parameters. This means that $X_1$ and $X_2$ conditioned on $(\Lambda_1,\Lambda_2)$ are independent Poisson processes with intensity functions 
 $\Lambda_1$ and $\Lambda_2$, respectively.
The (cross) pair correlation functions for this class of bivariate LGCP are isotropic, where the pair correlation function of $X_i$ is given by
\[
  g^\mathrm{iso}_{i}(r)=\exp\{\alpha_i^2 \exp(-r/\phi)+ \beta
  \exp(-r/\psi_i)\}, \qquad i=1,2,
\]
and the cross pair correlation function of $(X_1,X_2)$ is given by
\[
  c^\mathrm{iso}(r)=\exp\{\alpha_1 \alpha_2 \exp(-r/\phi)\}.
\]
Note that $c^\mathrm{iso}<1$ if $\alpha_1\alpha_2<0$ (the case of segregation between $X_1$ and $X_2$),
and $c^\mathrm{iso} >1$ if $\alpha_1\alpha_2>0$ (the case of co-clustering between $X_1$ and $X_2$). For the segregated processes, we chose $\alpha_1=-\alpha_2=1$, $\phi = .03$, $\beta = .25$, 
$\psi_1=.02$, and $\psi_2=.01$. For the co-clustered case, we used $\alpha_1=\alpha_2
= 1$ and the other parameters as for the segregated case. With these choices, the cross
correlation functions become 
\[ c^\mathrm{iso}_\mathrm{segr}(r)=\exp\{-\exp(-r/.03)\} \]
for the segregation case and 
\[ c^\mathrm{iso}_\mathrm{cluster}(r)=\exp\{\exp(-r/.03)\}\]
for the co-clustered case. Finally, we adjusted $\mu_1$ and $\mu_2$ so that the expected number of points after independent thinning is 200 or 400.

For the global estimator of $K_{12}$, we consider again the isotropic estimator \eqref{e:K12hat-global-iso},
since in each case the cross pair correlation function is isotropic, and estimation
of $\gamma_{12}^\mathrm{iso}(r)$ is less computationally intensive than that of
$\gamma_{12}(h)$. For the local estimator we consider the estimator \eqref{e:K12hat-local},
with $\rho_i$ estimated by the leave-out kernel estimator $\bar \rho$ from
\eqref{e:leaveoutkernel}. Similar to the $\{L(r) - r\}$-function used above, we
transform the $K_{12}$-function estimators into estimators of the
$\{L_{12}(r) - r\}$-function, by the one-to-one transformation
\[ L_{12}(r) - r = \sqrt{K_{12}(r)/\pi} - r. \]

\begin{figure}
  \centering
  \begin{tabular}{ccc}
    \includegraphics[width=0.33\textwidth]{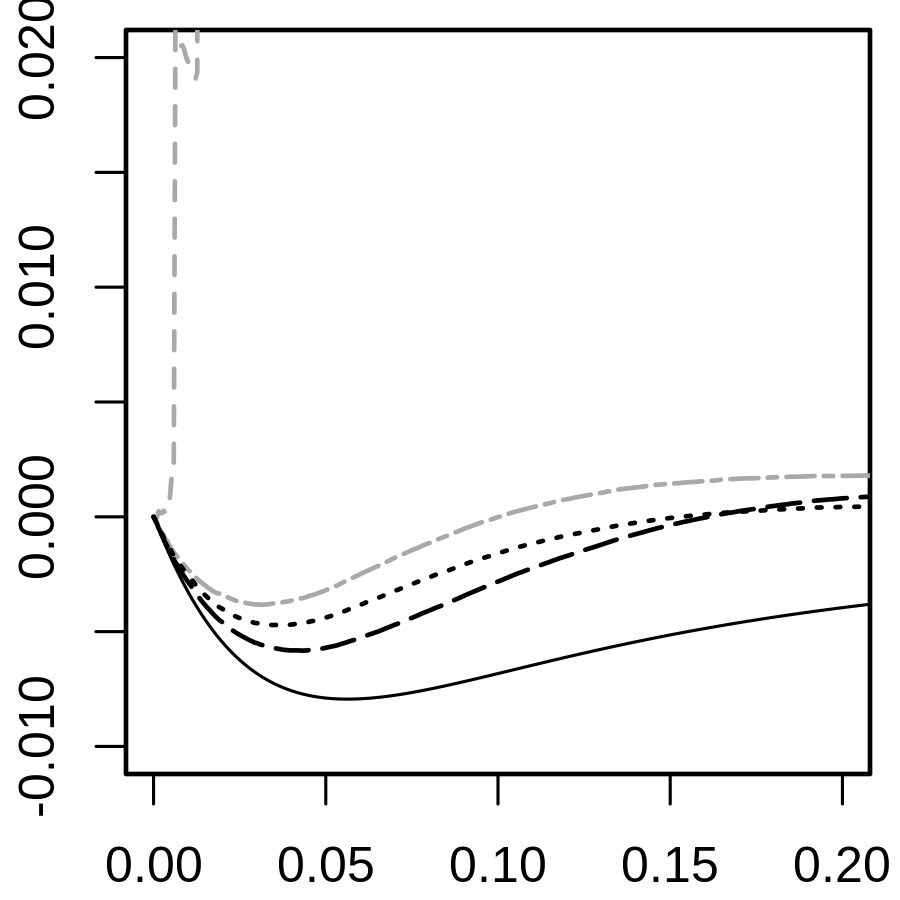}
    \includegraphics[width=0.33\textwidth]{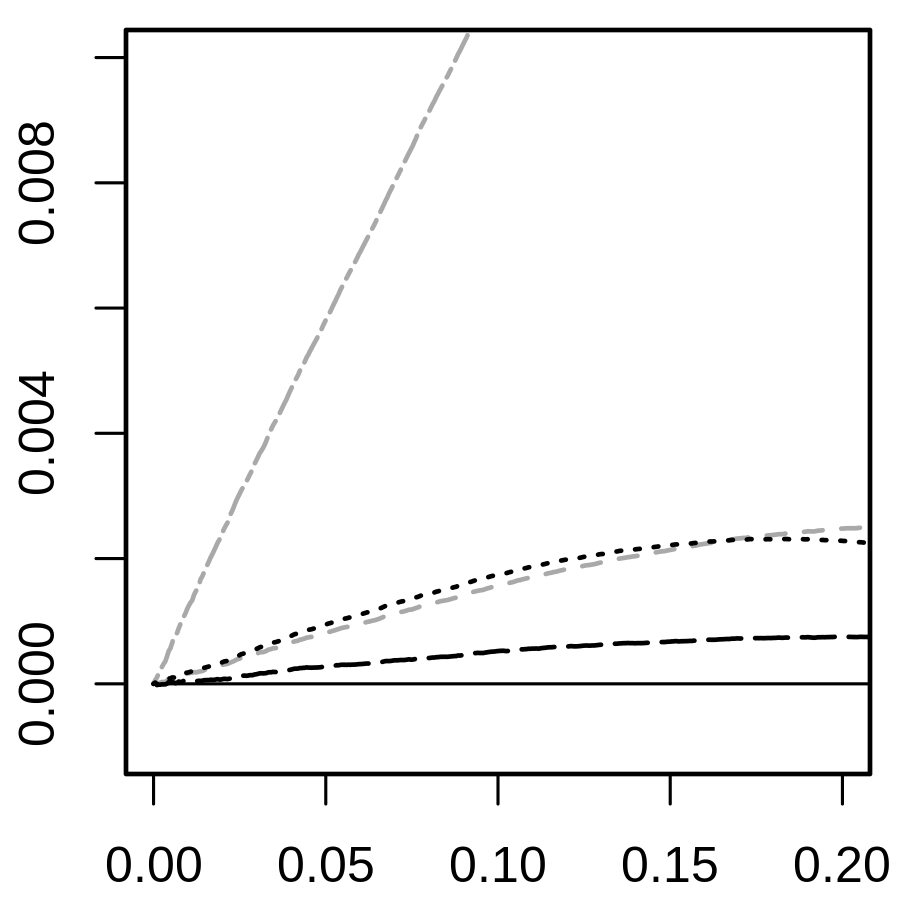}
    \includegraphics[width=0.33\textwidth]{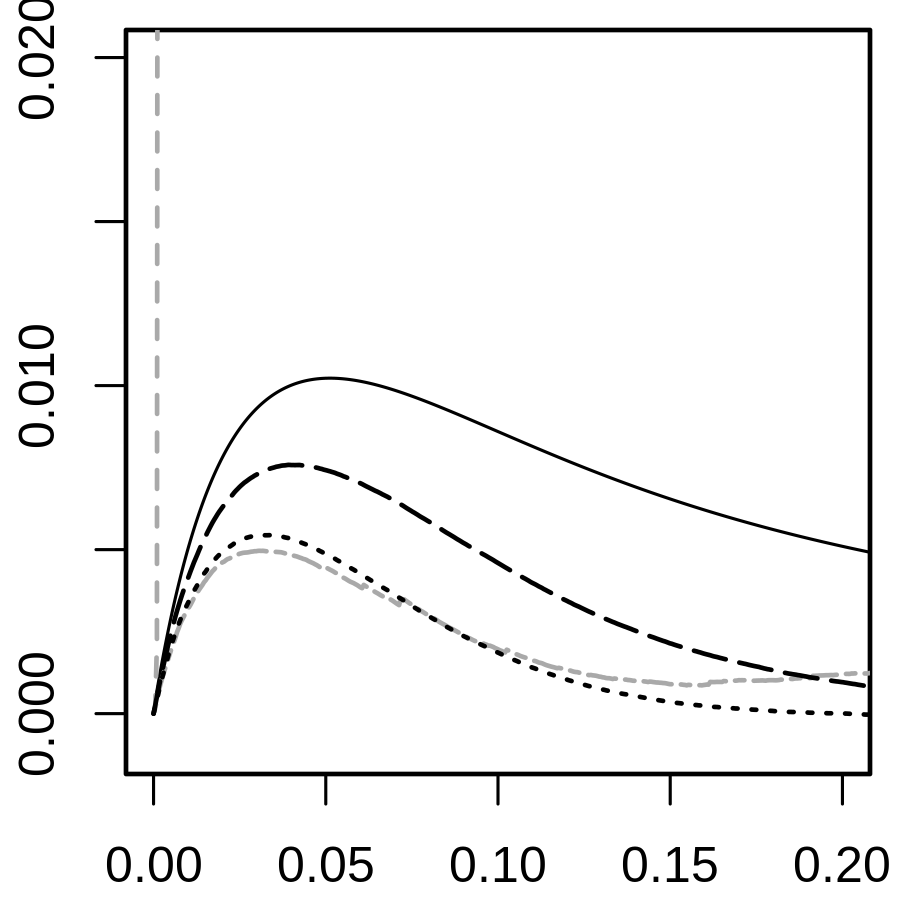}
    \end{tabular}
    \caption{Averages of estimates of cross-$L(r)-r$ in case of the waves
    intensity function with 400 simulated points on average. Left to right:
    segregation, independence, co-clustering. The estimators used are the
    standard global
    (\protect\includegraphics[width=1.2cm]{caption/t5w1-5b.png}~CVL,
    \protect \includegraphics[width=1.2cm]{caption/t3w1-5b.png}~LCV)
    and local
    (\protect\includegraphics[width=1.2cm]{caption/t6w1-5g.png}~CVL,
    \protect \includegraphics[width=1.2cm]{caption/t2w1-5g.png}~LCV)
    leave-out estimators of $K_{12}$ combined with the CVL
    and LCV methods for the bandwidth selection.
    True values of $L_{12}(r)-r$ are shown for comparison
    (\protect\includegraphics[width=1.2cm]{caption/t1w1b.png}).}
    \label{fig:bivariate}
  \end{figure}

Figure~\ref{fig:bivariate} shows averages of estimators of $L_{12}(r)-r$ in case of
the waves intensity and expected number of points equal to 400. The bandwidth
is selected using the CVL or LCV procedure applied to $X_1$. Table~\ref{t:bws-two} gives
selected bandwidth values for the pairs of spatial point processes we
considered. The results are similar to the one point process
  case. Both the segregated and co-clustered LGCP typically yield $\sigma_\mathrm{LCV} <
\sigma_\mathrm{CVL}$ while  the opposite is true for the Poisson
case. Further, the local estimators
are strongly biased, and the bias increases as the bandwidth $\sigma$
decreases: in the case of segregation and co-clustering, the local estimators are
better with CVL, while LCV is better in the case of independence. Note also that the
negative bias that is observed at small distances $r$ for $\hat
K_\mathrm{local}$ is absent here as predicted in the discussion in
Section~\ref{s:local-bias}. The bias for the global estimator with CVL is smaller than for the best
local estimators in each case.

\begin{table}
\caption{ Mean ($\pm$ st. dev.) of CVL and LCV selected bandwidths for the
simulated two point process cases. Expected number of points is 400 for each listed process.}
\label{t:bws-two}
\centering
\begin{tabular}{cccc}
Interaction type & Intensity function & $\sigma_\text{CVL}$ & $\sigma_\text{LCV}$ \\
\hline
Segregated  &   constant  & 0.063 (0.008)  &  0.038 (0.006) \\
  &   hole  & 0.062 (0.009)  &  0.039 (0.008) \\
  &  waves  & 0.064 (0.010)  &  0.040 (0.008) \\
Poisson  &   constant  & 0.048 (0.006)  &  0.60 (0.19) \\
  &   hole  & 0.048 (0.006)  &  0.28 (0.22) \\
  &  waves  & 0.051 (0.006)  &  0.19 (0.20) \\
Co-clustered  &   constant  & 0.062 (0.008)  &  0.040 (0.008) \\
  &   hole  & 0.060 (0.009)  &  0.040 (0.007) \\
  &  waves  & 0.064 (0.011)  &  0.040 (0.009) \\
\end{tabular}
\end{table}

\begin{figure}
  \centering
  \begin{tabular}{ccc}
    \includegraphics[width=0.33\textwidth]{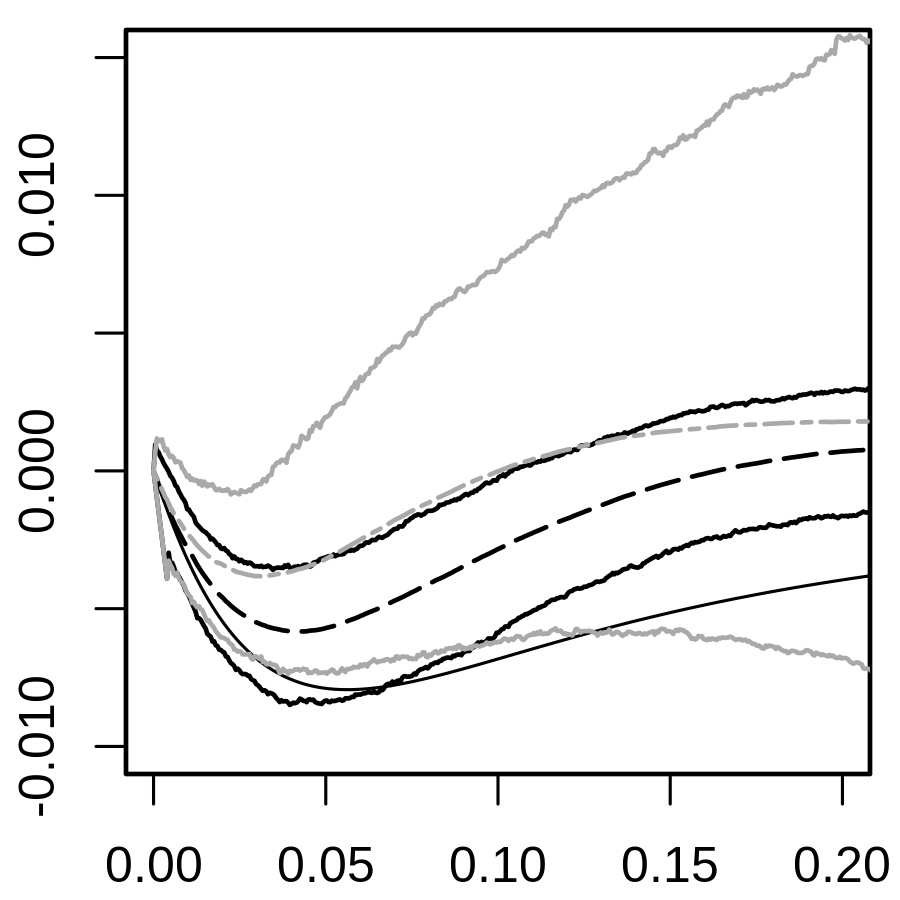}
    \includegraphics[width=0.33\textwidth]{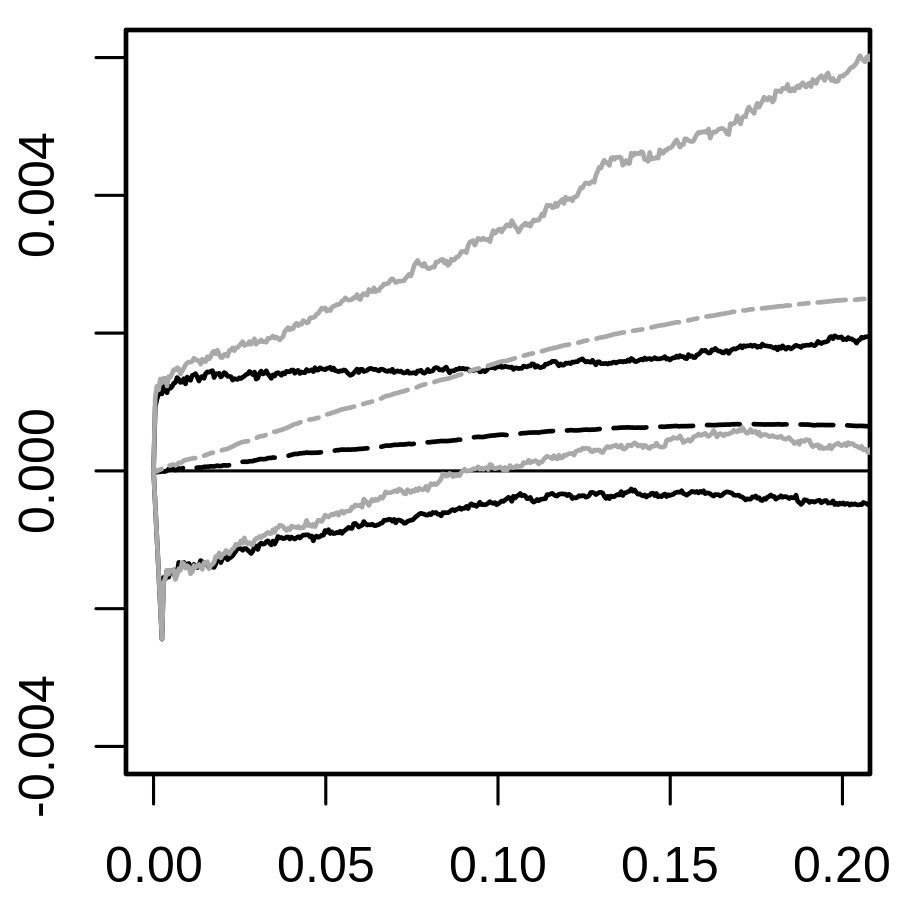}
    \includegraphics[width=0.33\textwidth]{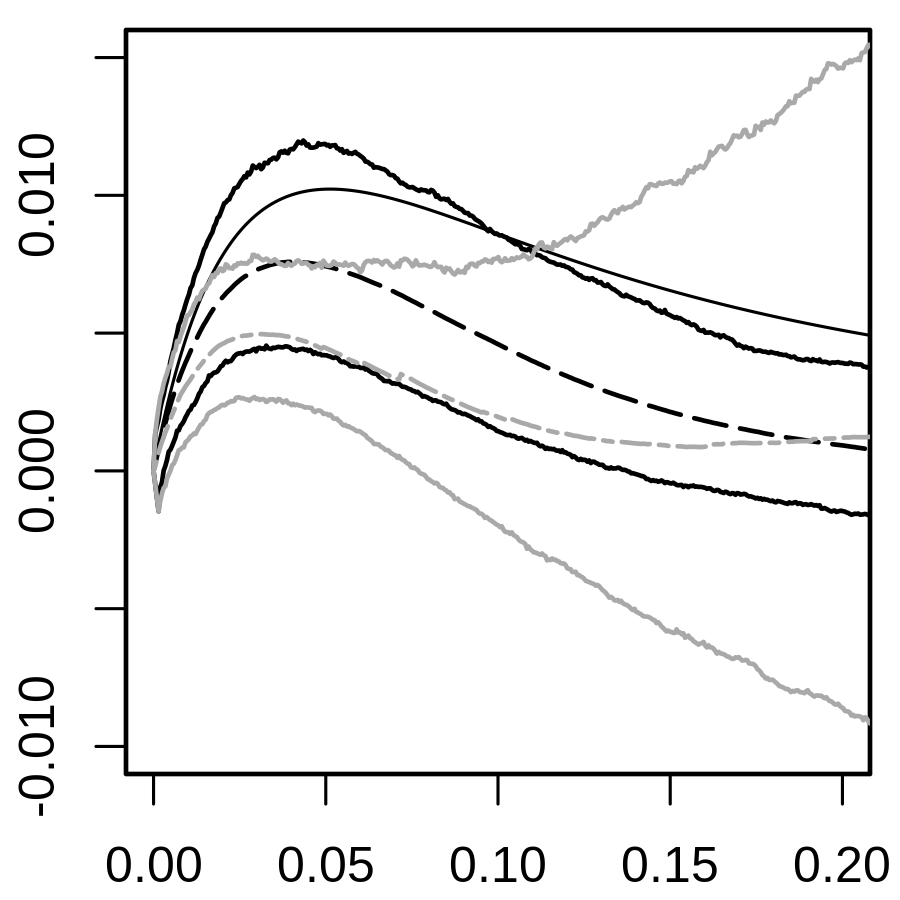}
    \end{tabular}
    \caption{Averages and 95\% pointwise probability intervals for estimates of $L_{12}(r)-r$
    in case of the waves intensity function with 400 simulated points on
    average. Left to right: segregation, independence, co-clustering. The
    estimators used are the leave-out global estimator
    (\protect\includegraphics[width=1.2cm]{caption/t5w1-5b.png})
    and the leave-out local estimator 
    (\protect\includegraphics[width=1.2cm]{caption/t6w1-5g.png}), with pointwise probability
    intervals shown in like shade. In each case, the bandwidth selection method
    was chosen to produce the least bias: LCV for the local estimator on the independent
    process, and CVL for all the other cases. True values of $L_{12}(r) - r$ are also shown
    (\protect\includegraphics[width=1.2cm]{caption/t1w1b.png}).
    }\label{f:var-two}
\end{figure}

To compare sampling variability for the estimators of the cross $K$-function, we
show pointwise 95\% probability intervals for estimated $L_{12}(r)-r$ in
Figure~\ref{f:var-two}. The bandwidth selection method that produces the least
bias in each case is shown. Table~\ref{t:imse-two} shows root integrated mean
square error of the estimators of $K_{12}$. In every case, the best global
estimator has smaller integrated mean square error than the best local
estimator, as expected from the considerations of Section~\ref{s:comparevar}.

\begin{table}

\caption{Root integrated mean squared errors $\times 10^2$ of local and global
$K_{12}$-function estimators with CVL and LCV bandwidths.}\label{t:imse-two}

\centering

\begin{tabular}{cccccc}
&&\multicolumn{2}{c}{$\hat K_{12,\text{local}}$} & \multicolumn{2}{c}{$\hat K_{12,\text{global}}$} \\
Interaction type & Intensity function &  CVL & LCV & CVL & LCV \\
\hline
 Segregated  &  flat & 0.65 & 390.125 & 0.161 & 0.181 \\
             &  hole & 0.69 &   4.574 & 0.171 & 0.185 \\
             & waves & 0.64 & 270.633 & 0.208 & 0.201 \\
 Independent &  flat & 1.03 &   0.066 & 0.024 & 0.049 \\
             &  hole & 1.09 &   0.112 & 0.026 & 0.109 \\
             & waves & 0.95 &   0.191 & 0.037 & 0.104 \\
 Co-clustered & flat & 0.92 &  18.783 & 0.234 & 0.262 \\
             &  hole & 0.97 &   3.510 & 0.239 & 0.265 \\
             & waves & 0.92 &   5.238 & 0.195 & 0.244 \\
\end{tabular}

\end{table}

\begin{figure}
  \centering
  \begin{tabular}{ccc}
    \includegraphics[width=0.33\textwidth]{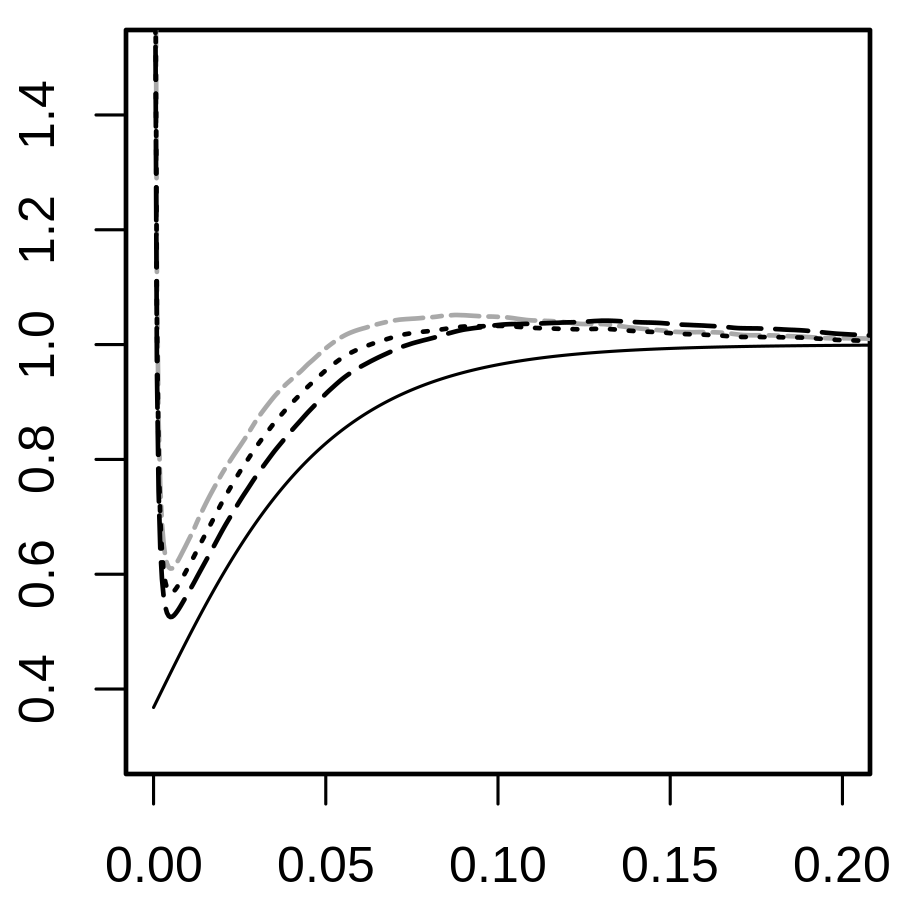}
    \includegraphics[width=0.33\textwidth]{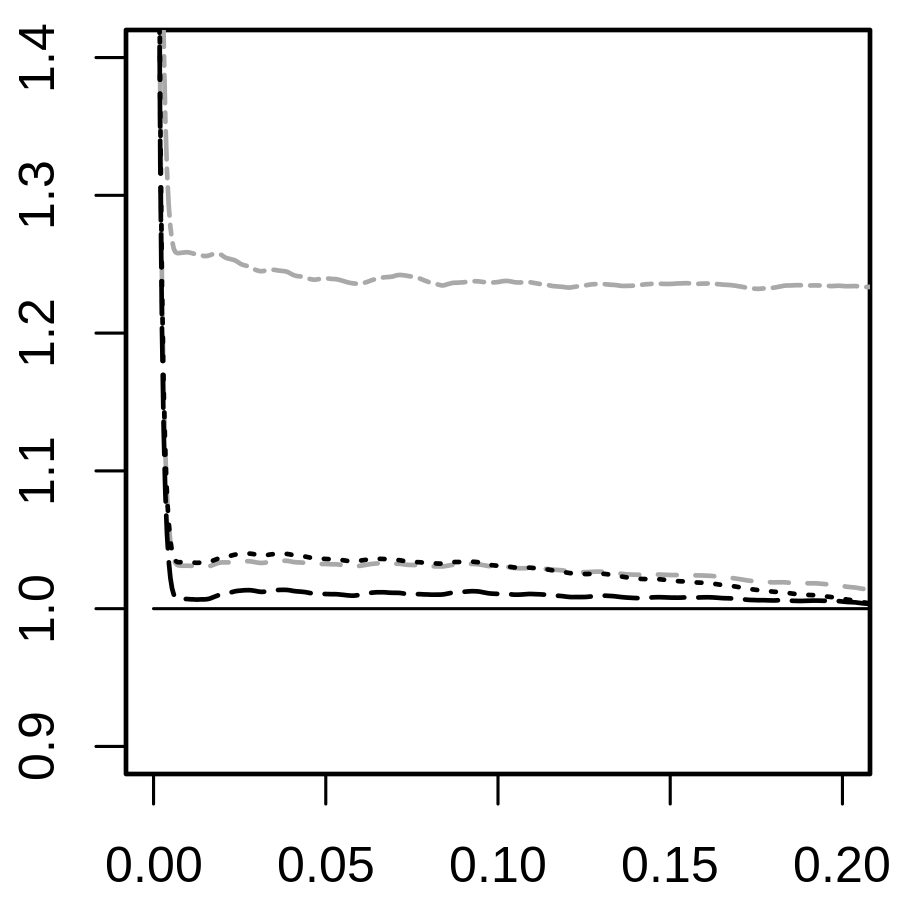}
    \includegraphics[width=0.33\textwidth]{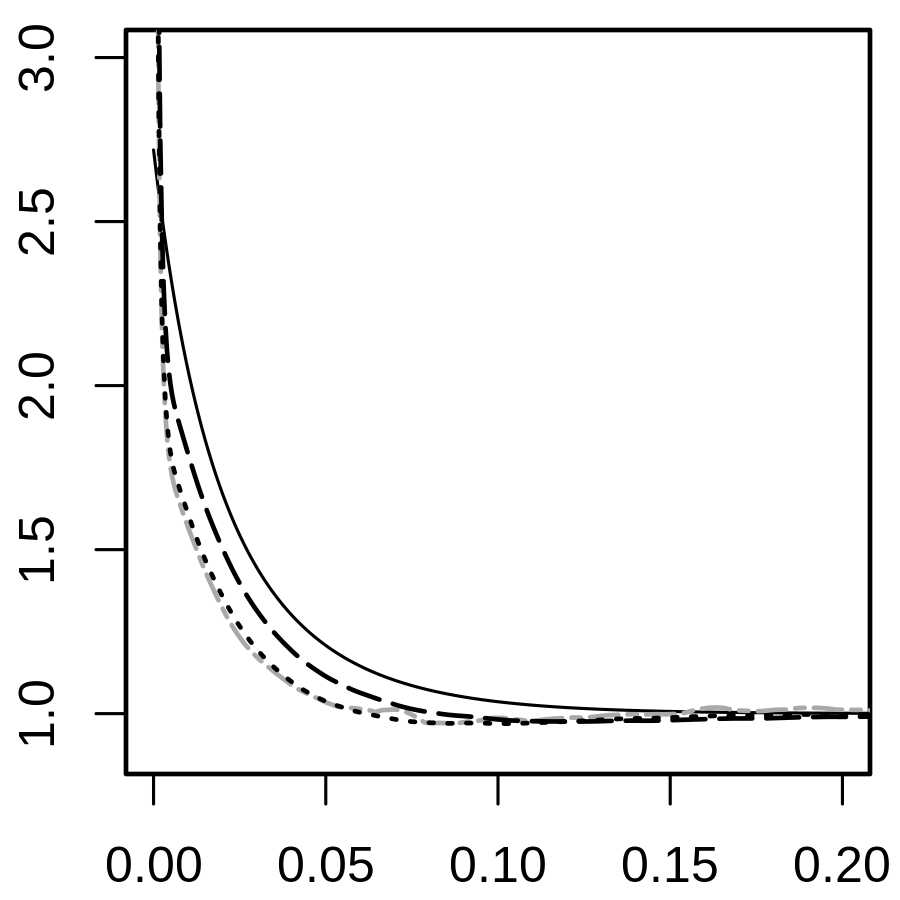}
    \end{tabular}
    \caption{Averages of estimates of $c(r)$ in case of the waves intensity function with 400 simulated points on average. Left to right: segregation, independence, co-clustering. The estimators used are the leave-out global
    (\protect\includegraphics[width=1.2cm]{caption/t5w1-5b.png}~CVL,
    \protect \includegraphics[width=1.2cm]{caption/t3w1-5b.png}~LCV)
    and local
    (\protect\includegraphics[width=1.2cm]{caption/t6w1-5g.png}~CVL,
    \protect \includegraphics[width=1.2cm]{caption/t2w1-5g.png}~LCV)
    estimators combined with the CVL and LCV methods for bandwidth selection.
    True values of $L_{12}(r)-r$ are shown for comparison
    (\protect\includegraphics[width=1.2cm]{caption/t1w1b.png}).} \label{fig:bivariatec} \end{figure}

  For the estimation of the cross pair correlation functions, the conclusions are
similar to those for the cross $K$-functions, see
Figure~\ref{fig:bivariatec}. The average of the global estimator is quite
close to the true cross pair correlation function, while the local estimator is
strongly biased. Note that $\hat c_\mathrm{local}^\mathrm{LCV}$ is missing for the
segregated and co-clustered processes, because the average values of that estimator
were extremely large. 

\subsection{Estimation of $K$-function using a parametric estimate for $\rho$}

Returning to the setting of a single point process $X$ as in the beginning of Section~\ref{s:sim}, we also consider the case of a parametric model where the intensity $\alpha>0$ of the underlying stationary point process (that is, before thinning) is unknown but
 the retention probability function $p$ that was used to thin the point process is known. Then a 
simple parametric estimator for $\rho$ is given by
\begin{equation}
\hat\rho_p(x) = N p(x)\big/\int_W p(x) \dd x, \label{e:par-rho}
\end{equation}
where
$N$ is the number of points in $X\cap W$. We apply this intensity estimator to
$\hat K_\text{local}$ and $\hat K_\text{global}$ for $1000$ realizations of
each interaction type, with the `waves' intensity function
and expected number of points equal to 400. In addition, we
  generate $1000$
simulations for each interaction type with a new thinning profile, `deep waves', given by
\[p_\text{deep}(x,y) = 1 - .9 \cos^2(5x),\qquad (x,y)\in[0,1]^2. \]
The deep waves profile is similar to the waves profile, but with much more extreme
intensity variations.

\begin{figure}
  \centering
  \begin{tabular}{ccc}
    \includegraphics[width=0.33\textwidth]{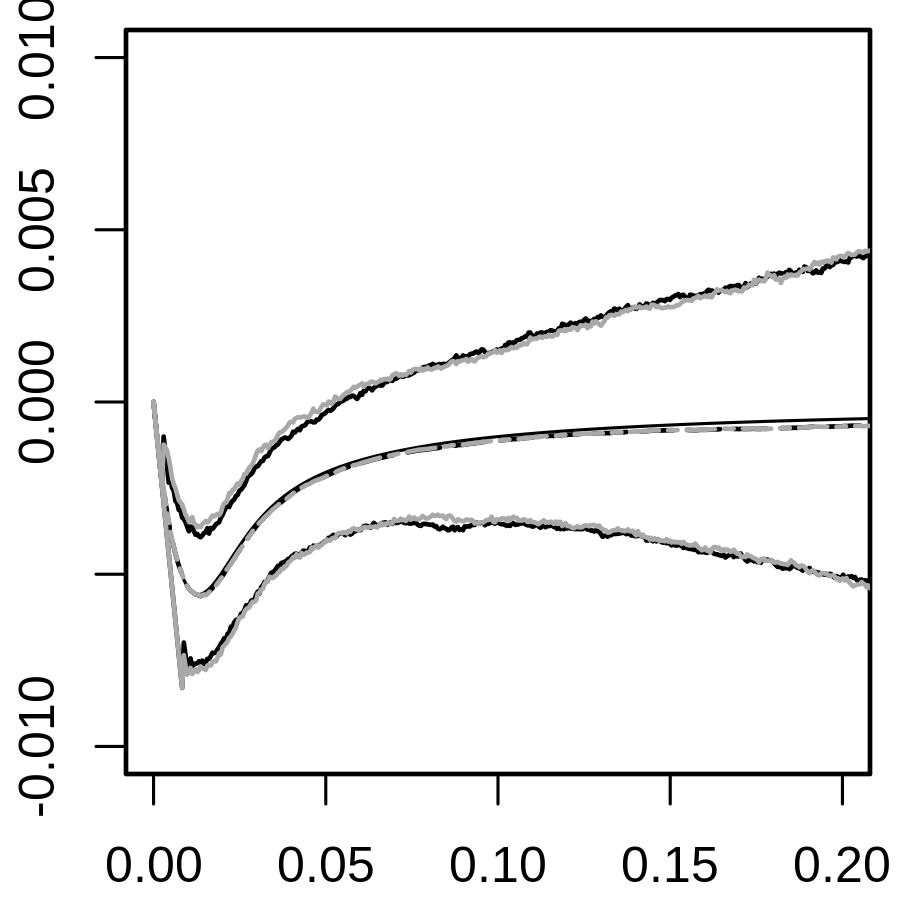}
    \includegraphics[width=0.33\textwidth]{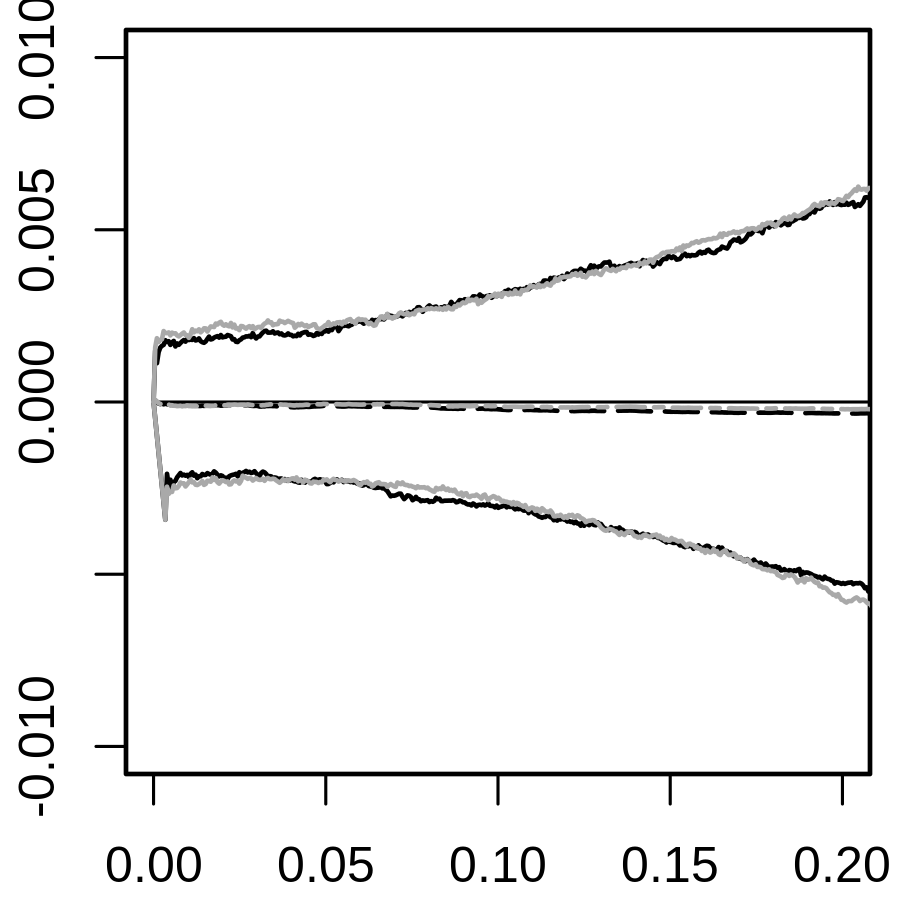}
    \includegraphics[width=0.33\textwidth]{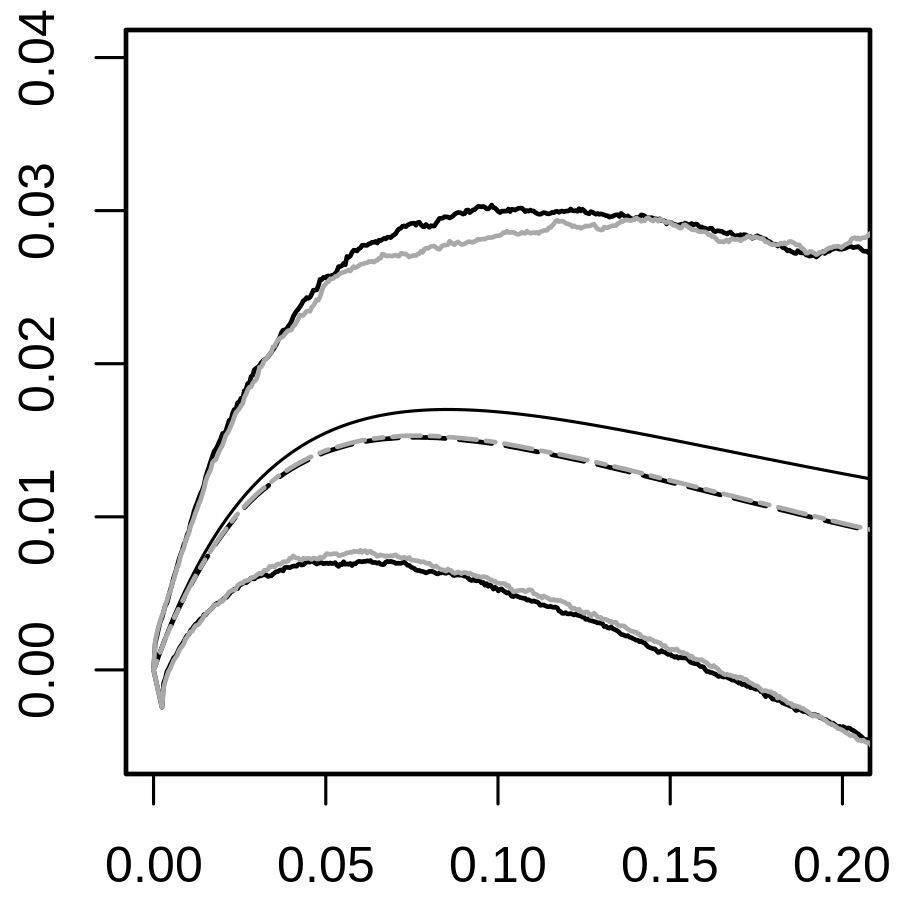}
    \\\includegraphics[width=0.33\textwidth]{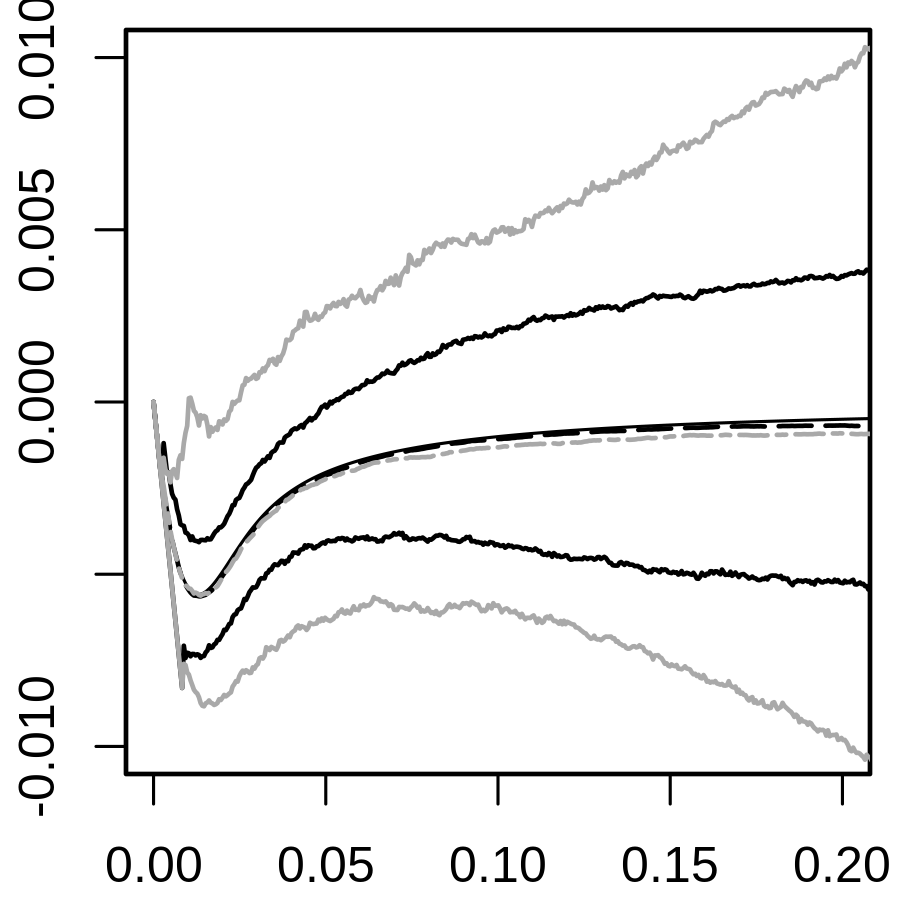}
    \includegraphics[width=0.33\textwidth]{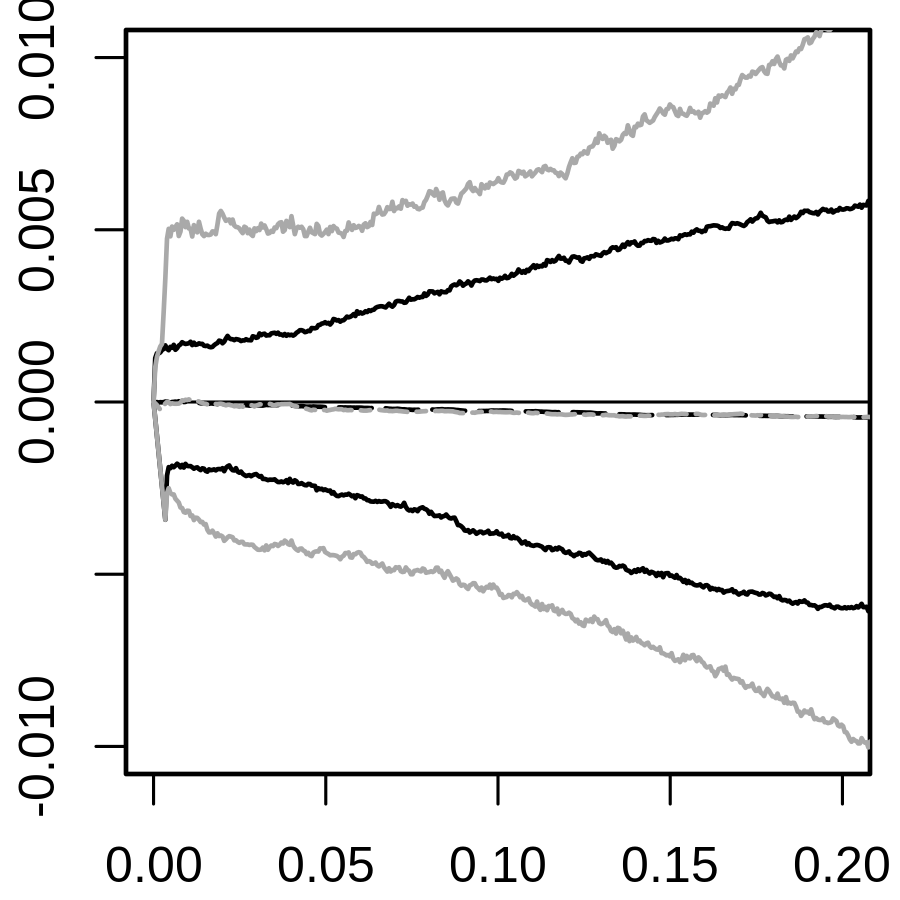}
    \includegraphics[width=0.33\textwidth]{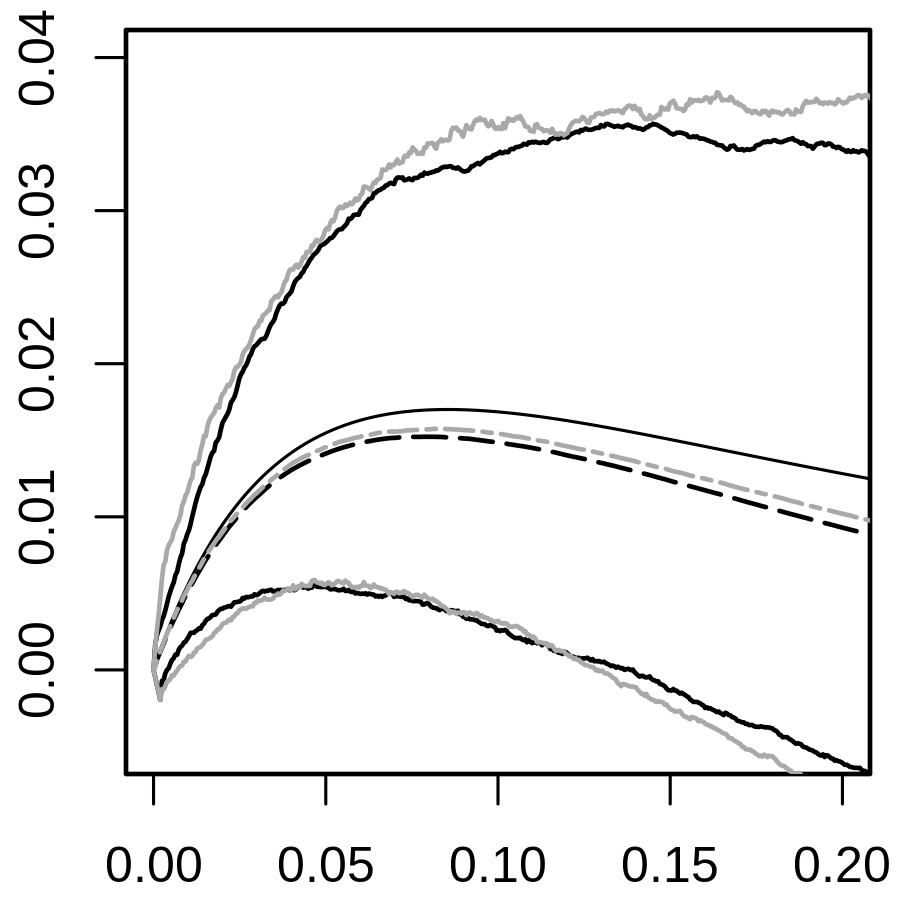}
    \end{tabular}
    \caption{Averages and 95\% pointwise probability intervals for estimates of
    $L(r)-r$ in case of the `waves' (top row) or `deep waves' (bottom row)
    intensity function with 400 simulated points on average. Left to right:
    DPP, Poisson, LGCP. The estimators used are the
    global
    (\protect\includegraphics[width=1.2cm]{caption/t5w1-5b.png})
    and local
    (\protect\includegraphics[width=1.2cm]{caption/t6w1-5g.png})
    estimators using the parametric intensity estimator \eqref{e:par-rho}.
    Pointwise probability intervals are shown in like shade. True values of
    $L(r) - r$ are also shown
    (\protect\includegraphics[width=1.2cm]{caption/t1w1b.png}).}
 \label{f:var-par}
\end{figure}

Pointwise probability intervals for estimates of $L(r)-r$ are shown in Figure~\ref{f:var-par},
and root integrated mean square error for estimates of $K$ are given in
Table~\ref{t:imse-par}. We observe that in all cases the error of the global
estimator is comparable to or better than the corresponding
local estimator. For the `waves' intensity function, the difference is small. Both estimators have larger error when applied to the patterns
with the `deep waves' intensity function. However, the performance of the
local estimator degrades much more strongly, reflecting the fact that regions of low intensity
are weighted more heavily in $\hat K_\text{local}$ than in $\hat
K_\text{global}$, as discussed in Section~\ref{s:comparevar}.  The LGCP yielded
the largest errors with the parametric intensity estimates, similar to our
observations with the kernel-based intensity estimates. We also note
that for the DPP and the Poisson process, using the parametric estimates for the `waves' intensity function results in higher integrated mean
square error than for the kernel-based estimates (Table~\ref{t:imse-one}). We
believe this is because the kernel-based estimates of $\rho$ are adapted to the random
local fluctuations of the point processes, similar to how homogeneous $K$-function estimates
have lower variance when using estimated intensity than true
intensity. However, for the LGCP, best results are obtained with
  the parametric estimates, which presumably are less prone to
  confounding of random clustering with variations in the intensity function.

\begin{table}
\caption{Root integrated mean squared errors $\times 10^2$ of local and global
$K$-function estimators with parametric intensity estimator, applied to point
processes with intensity function `waves' or `deep waves'.}\label{t:imse-par}
\centering
\begin{tabular}{cccc}
Interaction type & Intensity function &  $\hat K_\text{local}$ & $\hat K_\text{global}$ \\
\hline
    DPP & waves & .111 & .102 \\
        & deep waves & .227 & .103 \\
Poisson & waves & .132 & .122 \\
        & deep waves & .239 & .133 \\
   LGCP & waves & .416 & .417 \\
        & deep waves & .601 & .516 \\
\end{tabular}
\end{table}

\section{Extensions}\label{s:conrem}
The same sort of analysis as in Sections~\ref{s:K}-\ref{s:g} could be
applied to point processes defined on a non-empty manifold on which a group acts 
  transitively (a so-called homogeneous space), where the
   space is  equipped with a reference
  measure which is invariant under the group action. In this paper,
  the space was $\mathbb R^d$, the group action was given by
  translations, and the reference measure was Lebesgue
  measure. For example, instead we could consider the space to be a $d$-dimensional sphere,
  with the group action given by rotations and where the reference
  measure is the corresponding $d$-dimensional surface measure. Then the global and local estimators considered in this paper are simply modified to the case of the sphere by replacing Lebesgue with surface measure and using appropriate edge correction factors as defined in \cite{lawrence:etal:16}.
Similarly, our global
  estimators could also be extended to the case of spatio-temporal
point processes, as in \cite{GabrielSecondorder2009} and
\cite{MollerAspects2012}. 

\section{Conclusion}\label{s:conclusion}

According to our simulation studies, our new global estimators
outperform the existing local estimators in terms of bias and mean
integrated squared error when kernel or parametric estimators are
used for the intensity function. The kernel intensity function
estimators depend strongly on the choice of bandwidth and we
considered two different data-driven approaches, CVL and LCV, to bandwidth selection. In our simulation studies the two approaches gave
similar selected bandwidths in the LGCP case but very different results in case of
Poisson and DPP. This has a considerable impact on the $K$- and pair
correlation function estimators but the global estimators appear to
be much less sensitive to the choice of bandwidth selection method
than the local estimators.
The simulation studies with parametric estimates of the intensity function,
along with the theory of Section~\ref{s:comparevar}, indicate that the global
estimators are also much less sensitive to regions of especially low
intensity.
The improved statistical efficiency comes at a considerable extra
computational cost. Therefore, we especially recommend the global
estimators for situations where intensity variations are large  and where computational speed is not a
primary concern. 

\bibliographystyle{anzsj}

\bibliography{refAustNZJStat}

\end{document}